\documentclass[a4paper,12pt]{article}
\usepackage[english]{babel}
\usepackage{amsmath, amssymb,graphicx}
\usepackage{jheppub}

\newcommand{\be}{\begin{equation}}
\newcommand{\ee}{\end{equation}}
\newcommand{\bea}{\begin{eqnarray}}
\newcommand{\eea}{\end{eqnarray}}
\newcommand{\nn}{\nonumber}

\newcommand{\pr}{\prime}

\usepackage[normalem]{ulem}
\usepackage{color}

\title{
Waking and Scrambling in  Holographic Heating up}
\author[a]{D.S. Ageev}
\author[a]{and I.Ya. Aref'eva}
\affiliation[a]{Steklov Mathematical Institute, Russian Academy of Sciences, Gubkin str. 8, 119991
Moscow, Russia}

\emailAdd{ageev@mi.ras.ru}
\emailAdd{arefeva@mi.ras.ru}

\abstract{ We consider a holographic model of the heating up process. 
As a dual background  we take a geometry describing thin shell accretion on a black brane.   
We find explicitly  the time evolution  of the  mutual information during 
the non-equlibrium heating process from the initial temperature  $T_i$ to the final temperature  $T_f$ 
for the system of two  intervals in the 1+1 dimensional case. We calculate widths and 
separation of two
intervals for which 
 the time dependence of the mutual information has the bell-like form, 
 i.e.  it starts from zero value at the wake up time, then reaches a maximal value and  
 vanishes at the scrambling time. 
This form of the mutual information evolution was previously  found in photosynthesis. 
The zone of the bell-like configurations exists for  small distances
$x<\log 2/2\pi T_i$ only for the particular  interval  sizes. For $x$ large enough, i.e. $x>>\log 2/2\pi T_i$,  it
exists only for large enough interval  sizes and this zone
becomes  more narrow when
$T_i$  increases and becomes larger with increasing of   $T_f$.

 $$\,$$}


\begin{document}
\maketitle
\newpage

%\tableofcontents

\newpage

\section{Introduction}
 The AdS/CFT correspondence \cite{Malda,GKP,Witten} initially started as a duality between certain SYM theories and superstrings is a powerful tool for a description of general class of strongly-coupled quantum  theories. Broad phenomena  in QCD \cite{IA}-\cite{solana}, condensed matter \cite{Hartnoll:2009sz}, cosmology \cite{Easther:2011wh}  and even in biology  \cite{AV-photo} are described by the holographic duality.

One of the most challenging questions in traditional methods of quantum field theories is the thermalization process. In the AdS/CFT duality the temperature in the quantum field theory is  associated with the black hole (or black brane) temperature  in the dual background \cite{WittenTH}-\cite{Danielsson:1999fa}. The thermalization is described in the dual language by a black hole formation process,  see \cite{IA,DeWolf} and references therein. One of the simplest descriptions of this process is given by  the Vaidya deformation of a given background, see 
\cite{Balasubramanian:2011ur}-\cite{Aref'eva:2016dmy} and references therein.

   In this paper we investigate the holographic instantaneous heating process.  An instantaneous heating process can serve  as an  instant perturbation of  open quantum systems (see for example \cite{OV}), in particular, systems considered in biology (photosynthesis)  \cite{AV-photo,AVK,VK,bunch}. The dual description of this process  corresponds to a 
 massless shell  accretion 
 on   black hole   and  the corresponding metric describes the shell in the black hole background (we call it the BH-Vaidya metric).  
  In the context of
 photosynthesis the time evolution of the quantum mutual information that has the bell-like form during heating process is of special interest.  During  this evolution  the mutual information starts from zero value at the moment $t_{wup}$, then increases up to
 the maximum value and then vanishes   at the time $t_{scr}$.
  This type of evolution of 
 the mutual information has been found numerically in the holographic model of photosynthesis \cite{AV-photo}, where the thick shell BH-Vaidya metric has been considered for particular values of 
 parameters, specifying the geometry of the  system composed of two strips or intervals.    Our goal is to perform a detailed study of the regions in the space of parameters, where the bell-like form of the time evolution of the 
mutual information takes place. Also we find the corresponding  wake up $t_{wup}$ and scrambling $t_{scr}$ times.  It is useful to have an explicit formula for the entanglement entropy evolution  instead of numerical results for the thick shells
 \cite{CVJ,Keranen,AT-mut,Aref'eva:2016dmy,Ageev:2016viy} for this purpose. To this end 
we  generalize the explicit results  obtained for  the thin AdS-Vadya shell \cite{Balasubramanian:2011ur,Lopez,ABK,Liu} to  
the thin BH-Vaidya shell  model.  The explicit formulae for the time evolution of the entanglement entropy  for the thin BH-Vaidya shell  are more cumbersome then ones for the thin AdS-Vadya shell, however they allow us to simplify the   detailed analysis  of the mutual information evolution \cite{bala-mut,AT-mut,Ali}, \cite{AV-photo},  in particular, find 
 the wake-up and scrambling times.

The bell-type behaviour of the time evolution of the mutual information takes place in a part of a larger region of parameters in which 
the finite  scrambling time exists, but there is no the wake up regime.  In other words, the bell zone of parameters is a part of the scrambling zone.
The scrambling zone in the space of geometrical parameters of our model depends on the  final temperature only, meanwhile the bell
zone depends on the initial and final temperature.

 This paper is organized as following. 
In Sect.2 we describe the dual geometry of the  heating process and present the explicit formulae 
for entanglement entropy evolution.
Sect.3 is devoted to the study of  the evolution of  the mutual information during the heating process.
Here we estimate the sizes of scrambling and bell zones in the parametric space as well as the scrambling  
and wake up  time dependences on the final and initial temperatures.

\section{BH-Vaidya geometry and geodesics}
\subsection{BH-Vaidya geometry}
In this paper we consider black hole collapsing from an initial state defined by horizon position $z_H$ to the final state with horizon $z_{h}$ as a dual background. The BH-Vaidya metric is given by

\bea\label{Vm1}
\label{Vm1}
v<0:&\,\,\,&
ds^2=\frac{1}{z^2}\left( -f_{H}(z)dt^2+ \frac{dz^2}{f_{H}(z)}+dx^2\right),\,\,\,\,v=t-z_H \text{arctanh} \frac{z}{z_H},\\
\label{Vm2}v>0:&\,\,\,&
ds^2=\frac{1}{z^2}\left( -f_{h}(z)dt^2+ \frac{dz^2}{f_{h}(z)}+dx^2\right),\,\,\,\,\,\,v=t-z_h \text{arctanh} \frac{z}{z_h},
\eea
and the functions  $f_{H}$ and $f_{h}$ are defined as
\bea\label{BHpm}
f_H=1-\left(\frac{z}{z_{H}}\right)^2 ,\,\,\,\,
f_{h}=1-\left(\frac{z}{z_{h}}\right)^2,\,\,\,\,z_h<z_{H},
\eea
The initial and finite temperatures are
\be
\label{temperatures}
T_i=\frac{1}{2\pi z_H},\,\,\,\,\,\,T_f=\frac{1}{2\pi z_h}\ee
In more compact notations the metric is
\bea
ds^2&=&\frac{1}{z^2}\left( -f(z,v)dt^2+ \frac{dz^2}{f(z,v)}+dx^2\right),
\eea
where
\bea
f(z,v)&=&\theta(v)f_{h}(z)+\theta(-v)f_H(z).
\eea
This metric describes the shell located at $v=0$. Note, that the case $z_H<z_h$ corresponds to a model of cooling \cite{IAIV} and this model violates NEC condition \cite{Liu}.

\subsection{The geodesics in the BH-Vaidya background.}
The action for a geodesic connecting two points on the boundary $-\ell/2$ and $\ell/2$ at the boundary time moment $t$ in the background  with metric \eqref{Vm1} and parametized as $z=z(x)$ is given by

\bea\label{action1}
S(\ell)&=&\int_{-\ell/2}^{\ell/2} dx \frac{\sqrt{Q}}{z},\,\,\,\,\,\,\,\,\,\,\,\,\,\,\,
Q=1-2v'z'-f(z,v) v^{ \prime 2},
\eea
 where 
\bea\label{BC}
z(\pm \ell/2)=0,\,\,\,\,\,v(\pm \ell/2)= t.
\eea
The symmetry of the problem implies that $z'(0)=v'(0)=0$ and we denote:
\bea
z(0)=z_*,\,\,\,\,\,
v(0)=v_*.
\eea
In the regions $v>0$ and $v<0$ the  dynamical system \eqref{action1} has two  integrals of motion $ E_\pm$ and $
 {J}_\pm$, 
\bea
 {E}_\pm=z'+f v',\,\,\,\,\,
 {J}_\pm=z \sqrt{Q}.
\eea
where we  denote   values of these integrals  over the shell ($v>0$) as $E_+$, $J_+$ and under the shell ($v<0$) as $E_-$ and $J_-$.
The integrals of motion $J_{\pm}$ are  equal due to  the $x$-independence of  $Q$ and they are equal to the value of $z$ at the "top" point $z_*$, the point  where $z'=v'=0$, i.e.
\bea\label{sys1}
z_*=z_-\sqrt{1-2v_-^{\pr}z_{-}^{\pr}-f_h(z) v_-'^2}=z_+\sqrt{1-2v_+^{\pr}z_{+}^{\pr}-f_H(z) v_+'^2}.
\eea
Above the shell we have $E_-=0$ 
\bea\label{sys2}
z_-'+f_H (z)v_-'=0,
\eea
and when crossing the shell  $E$ is not conserved, $E_+\neq E_-$. In the crossing point we can write down the equality for $E_+$
\bea\label{sys3}
  z_{+c}'+f_H(z_c)v_{+c}'=z_{+}'+f_H(z)v_+'.
\eea
The geodesic passing through the shell $v=0$ at a point $z_c$ has  \textcolor{magenta}{a} discontinuity. 
Also, the integrals of motion $J_{\pm}$ can be matched in $z_c$ 
\bea\label{sys4}
z_-\sqrt{1-2v_-^{\pr}z_{-}^{\pr}-f_H(z) v_-'^2}\Big|_{z_c}&=&z_-\sqrt{1-2v_-^{\pr}z_{-}^{\pr}-f_H(z) v_-'^2},\\\nn
z_+\sqrt{1-2v_+^{\pr}z_{+}^{\pr}-f_h(z) v_+'^2}\Big|_{z_c}&=&z_+\sqrt{1-2v_+^{\pr}z_{+}^{\pr}-f_h(z) v_+'^2},
\eea
and from $\eqref{sys2}$ we have the condition at  the crossing point
\bea\label{sys5}
z_{-c}'+f_H(z_c)v_{-c}'&=&0.
\eea
Integrating the equations of motions across the shell one can derive the continuity condition for function $v$
\be\label{sys6}
v_{-c}'=v_{+c}'.
\ee
Solving system of equations \eqref{sys1}-\eqref{sys6} we can express   the integral of motion $E_+$ and derivatives $z_{\pm}^{\pr}$,$v_{\pm}^{\pr}$ in terms of $z_c$ and $z_*$. For example, $E_+$ is
\bea
E_+&=&-\frac{z_c \left( z_H^2-z_{h}^2\right)   }{2 z_{H} z_h^2   }\sqrt{\frac{z_*^2-z_c^2}{z_{H}^2-z_c^2}}.
\eea

Let us introduce the following   notations (compare with \cite{Balasubramanian:2011ur}):
\bea\label{list-var}
z_*&=&\frac{z_c}{s},\,\,\,\,\,\,
z_c=\frac{z_h}{\rho},\,\,\,\,\,\,
z_{H}=\frac{z_h}{\kappa},\,\,\,\,\,\,
c=\sqrt{1-s^2},\nn \,\,\,\,\,\,
\Delta=\sqrt{\rho^2-\kappa^2},\,\,\,\,\,\,
\eea

In  terms of these variables the form of $E_+$ is  
\bea
\label{E+}
E_+&=&-\frac{c (1-\kappa^2)}{2 \Delta  \rho  s}.
\eea

The expression for the boundary time $t$  in terms of $E_+$ is:
\bea\label{to}
t=\int_0^{z_c} \frac{dz}{f_h(z)}\left(\frac{E_+ z}{\sqrt{\left(z_*^2-z^2\right) f_h(z)+E_+^2 z^2}}-1\right),
\eea
Evaluating integral in the RHS of \eqref{to} and substituting here $E_+$ from \eqref{E+} we get

\be\label{time}
\coth \frac{t}{z_h}=\frac{\left(-\kappa ^2+2 \rho ^2+1\right) c+2
   \rho \Delta}{2 \left(\Delta+\rho  c\right)}.\\
\ee

It is straightforward to obtain the expression for the extension $\ell$ of geodesics along the boundary. 
It is the sum of two terms, corresponding to two  parts  of the geodesics, under and over the shell. Explicitly we have
$\ell/2=\ell_-+\ell_+$ , where $\ell_-$ and $\ell_+$ are
\bea\label{intl}
\ell_-/2&=&\int_{z_c}^{z_*} \frac{z^2 dz}{\sqrt{z^2 \left(z_*^2-z^2\right) f_{H}(z)}}\\\nn
\ell_+/2&=&\int_{0}^{z_c}\frac{z^2 dz}{\sqrt{z^2 \left(z_*^2-z^2\right) f_{h}(z)+E_+^2 z^4}},.
\eea
Introducing the notation $\gamma^2=1-\kappa^2 $ and integrating in \eqref{intl} we get the explicit formulae 
\bea\label{lexpl}
\frac{\ell_-}{2}&=&\frac{z_h}{2\kappa} \log  \left( \frac{(c \kappa +\Delta  s)^2}{\rho ^2 s^2-\kappa ^2} \right)\\\nn
\frac{\ell_+}{2}&=&\frac{z_h}{2}\log \left(\frac{c^2 \gamma ^4-4 \Delta  \left(\Delta  \left(\rho ^2-2\right) s^2+\Delta- c s \sqrt{4
    \Delta ^2 \left(\rho ^2-1\right)+
   \gamma ^4}
   \right)}{c^2 \gamma ^4-4 \Delta ^2 (\rho  s-1)^2}\right)
\eea
 The expressions for the length of each part of the geodesic
\bea\label{S-int}
{\cal L}_-/2&=&z_*\int_{z_c}^{z_*}\frac{dz}{z^2\sqrt{\left(1-z^2/z_{H}^2\right)\left(z_*^2/z^2 -1\right)}}\\
 {\cal L}_{+, \varepsilon }/2&=&z_*\int_{\varepsilon}^{z_c}\frac{dz}{z^2\sqrt{\left(1-z^2/z_{h}^2\right)\left(z_*^2/z^2-1\right)+E_+^2}}
\eea 
 where we introduced regularization $\varepsilon$ for divergent piece of the geodesic $ {\cal L}_{\varepsilon}$.
 
 After integration and standard removal  of the divergent part coming from $\cal{L_{+,\varepsilon}}$ when $\varepsilon$ goes 
 to zero, we get
 \bea\label{LL}
&&    {\cal  L}_-/2=\frac{1}{2}\log \left(\frac{\rho  \left(c^2 \rho +2 c
   \Delta+\rho \right)-\kappa ^2}{\Delta^2-c^2\rho^2}\right),\\\nn 
&&   \nn {\cal  L}_+/2=\log \left(\frac{2 \Delta  z_h}{\sqrt{{\cal K}_{-} {\cal K}_{+}}}\right),\\\nn
   &&{\cal K}_{-}=2
   \Delta  (\rho -1)-c \left(\kappa ^2-2 \rho  (\rho -1)-1\right),\\&&
   \nn {\cal K}_{+}=2
   \Delta  (\rho +1)-c \left(\kappa ^2-2 \rho  (\rho +1)-1\right).
\eea   
  Formulae \eqref{time},\eqref{lexpl} and \eqref{LL} in the limit $\kappa \rightarrow 0$, $\Delta \rightarrow \rho$ and $\gamma \rightarrow 1$  reproduce the corresponding formulae for thermalization from \cite{Balasubramanian:2011ur}. The holographic entanglement entropy for a single interval  is given  by the length of  the geodesic that is anchored on this  interval  \cite{RT}. We can solve the equation \eqref{time}  obtaining the time dependence analyticaly as some function $\rho=\rho(s,t)$ (we do not write out it explicitly due to its length). Substituting $\rho=\rho(s,t)$ in \eqref{list-var}-\eqref{LL}, we have
   \bea\label{EE}
S(s,t)&=&2\left(  {\cal L}_-(s,t) + {\cal L}_+(s,t)\right),\\\nn
\ell(s,t)&=&2(\ell_-(s,t)+\ell_+(s,t)).
   \eea

    Next we invert numerically the functions of one variable \eqref{EE} to get $S(\ell,t)$.

\section{Evolution of the mutual information}
\subsection{Typical evolution of the mutual information}
Now let us consider the mutual information evolution.
The mutual information for two intervals system of lengths $\ell_1$ and $\ell_2$ divided by distance $x$ at the time moment $t$ is defined as
\bea
I(\ell_1,\ell_2,x,t)=S(\ell_1,t)+S(\ell_2,t)-
\left(S(\ell_1+\ell_2+x,t)+S(x,t) \right).
\eea

The behaviour of the mutual information  thermalization has been investigated in many previous studies \cite{bala-mut,AT-mut}, where  the AdS-Vaidya model has been considered. We study the   general features of the mutual information behaviour in the  BH-Vaidya model,
 i.e. when the initial state is already thermal.
 A quantum quench in the quantum theory  starting from a thermal 
initial state has been studied in \cite{QQT}.
 
 % The local quench in the thermal 1+1 conformal systems has been studied in \cite{QQT}. 

We are looking for a very special  mutual information evolution, namely the evolution that has the form of the bell.
As has been noted in \cite{AV-photo},  this type of behaviour of the holographical mutual information reproduces results of numerical 
calculations
for  a special non-equilibrium open quantum system describing  the  photosynthesis  \cite{NUM}.

\begin{figure}[h!]
\centering\begin{picture}(185,150)
\put(0,0){\includegraphics[scale=0.5]{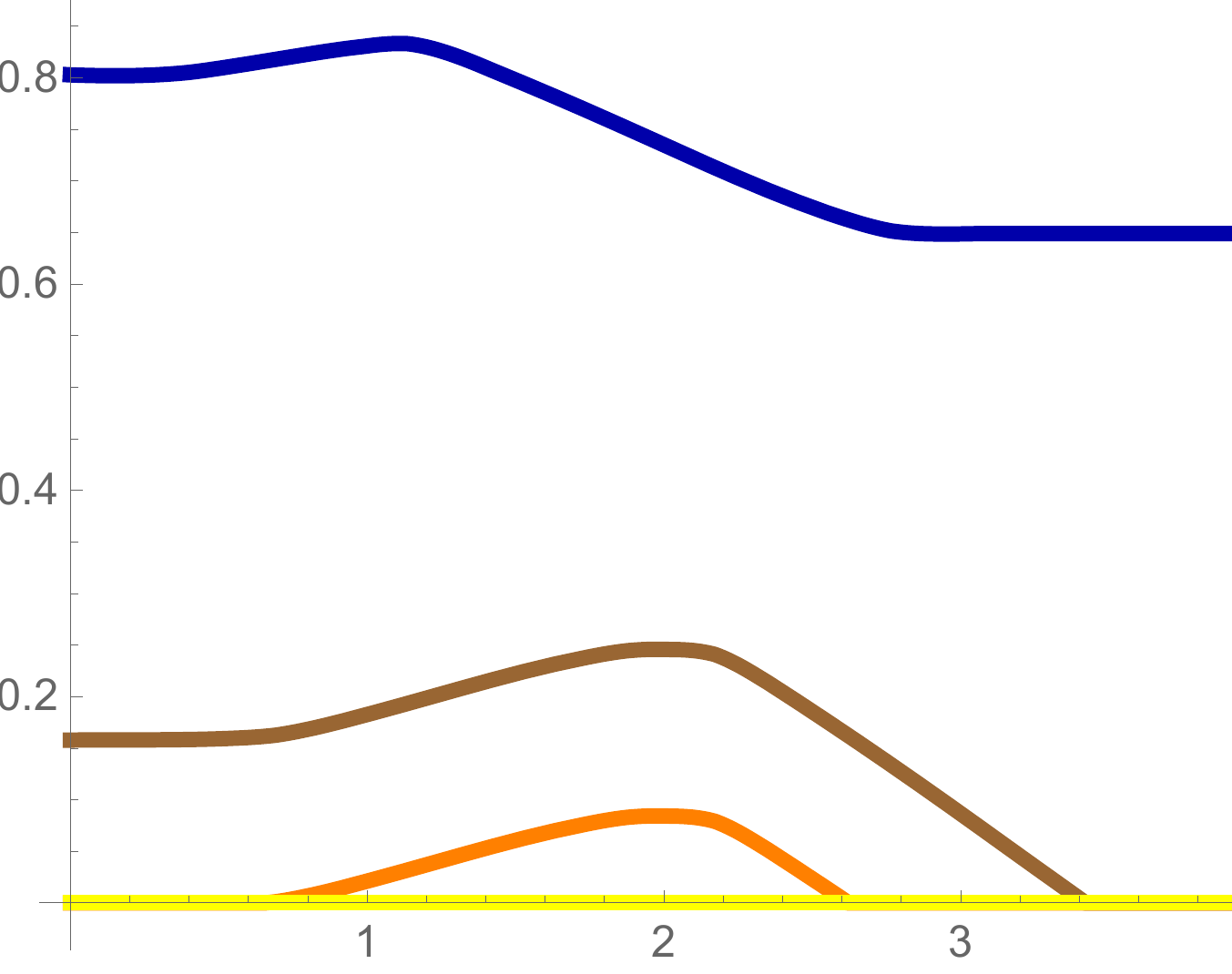}}
  \put(210,0){$t$} 
\put(10,149){\line(1,0){55}}
\put(10,117){\line(1,0){135}}
\put(10,53){\line(1,0){95}}
\put(10,26){\line(1,0){95}}
\put(-22,153){$I_{max}$}
\put(-28,140){$I_{start}$}
\put(-30,117){$I_{min}$}
\put(-40,23){$I_{max,bell}$}
\put(-40,54){$I_{max,scr}$}
\put(-40,36){$I_{start,scr}$}
 \put(35,-5){$t_{wup}$}
   \put(9,160){$I$}
  \put(130,-5){$t_{scr,\,bell}$}
    \put(175,-5){$t_{scr}$}
    \put(46,8){\line(0,1){8}}
     \put(133,8){\line(0,1){8}}
      \put(170,8){\line(0,1){8}}
\end{picture}\\
%$$\,$$
 \caption{Different regimes of the mutual information evolution in the heating process  of two disjoint intervals. The blue curve  starts from a non-zero value of the mutual information $I_{start}\neq0$, then increases and reaches  the maximum $I_{max}$, and then starts to decrease up to a fixed value $I_{min}\neq 0$. The brown curve corresponds to the scrambling behaviour,
  $I(t_{start,scr})\neq 0$,  $I(t_{scr})=0$.
The orange curve corresponds to the bell regime,  $I(t_{wup})=I(t_{scr,\,bell})=0$, $t_{wup}<t_{scr,\,bell}$.  The yellow line corresponds to mutual information vanishing during all process.}
 \label{fig:evol}
\end{figure}
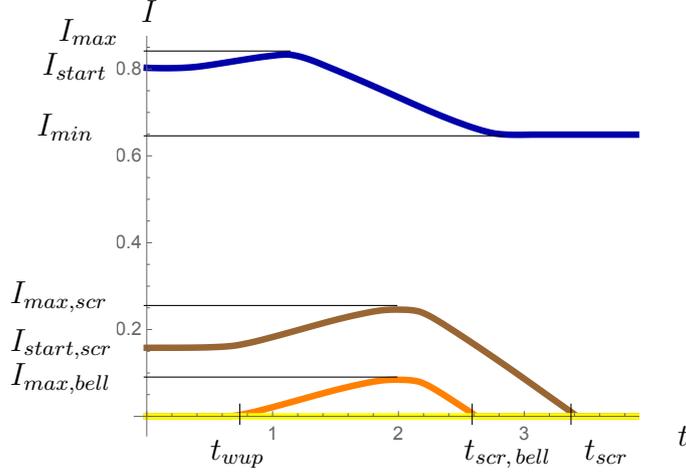

There are four different regimes of the mutual information time evolution during  the heating process  of two intervals:
\begin{itemize}
\item
the regime where the mutual information  starts from a non-zero value,  then increases and reaches  the maximum $I_{max}$. After that it starts to decrease up to a fixed  positive value $I_{min}\neq 0$ (the blue line in Fig.\ref{fig:evol}  shows this type of the time evolution of the mutual information);
\item the regime  with  a scrambling point, i.e. the regime, where at the time $t_{scr}$ the mutual information  vanishes
(the brown line in Fig.\ref{fig:evol});  
\item  the regime,  where
at the moment $t_{wup}$ the mutual information starts from zero value (the orange line in Fig.\ref{fig:evol});  
\item   the regime with the vanishing mutual information
(the yellow line in Fig.\ref{fig:evol}). 
\end{itemize}

These  regimes are similar to  the time evolution of the mutual information during  the 
thermalization  process \cite{bala-mut,AT-mut,Ali} and physical explanations  of these regimes are the same. 

\begin{figure}[tbp]
\centering
 \includegraphics[scale=0.3]{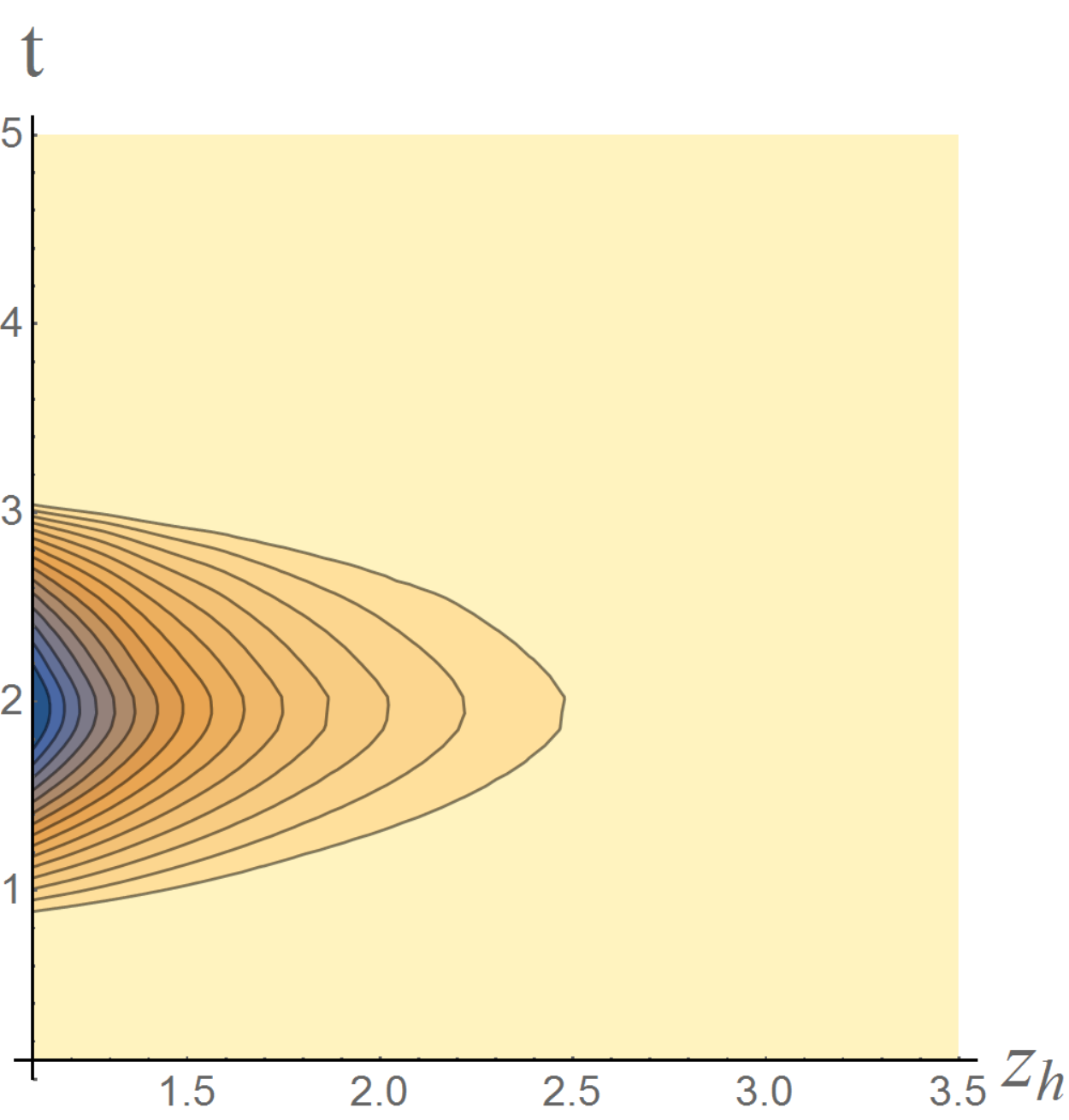} \,\,\,\,\,\,\,\,\,\,\,\, \,\,\,\,\,\, \,\,\,\,\,\, \,\,\,\,\,\,  
  \includegraphics[scale=0.3]{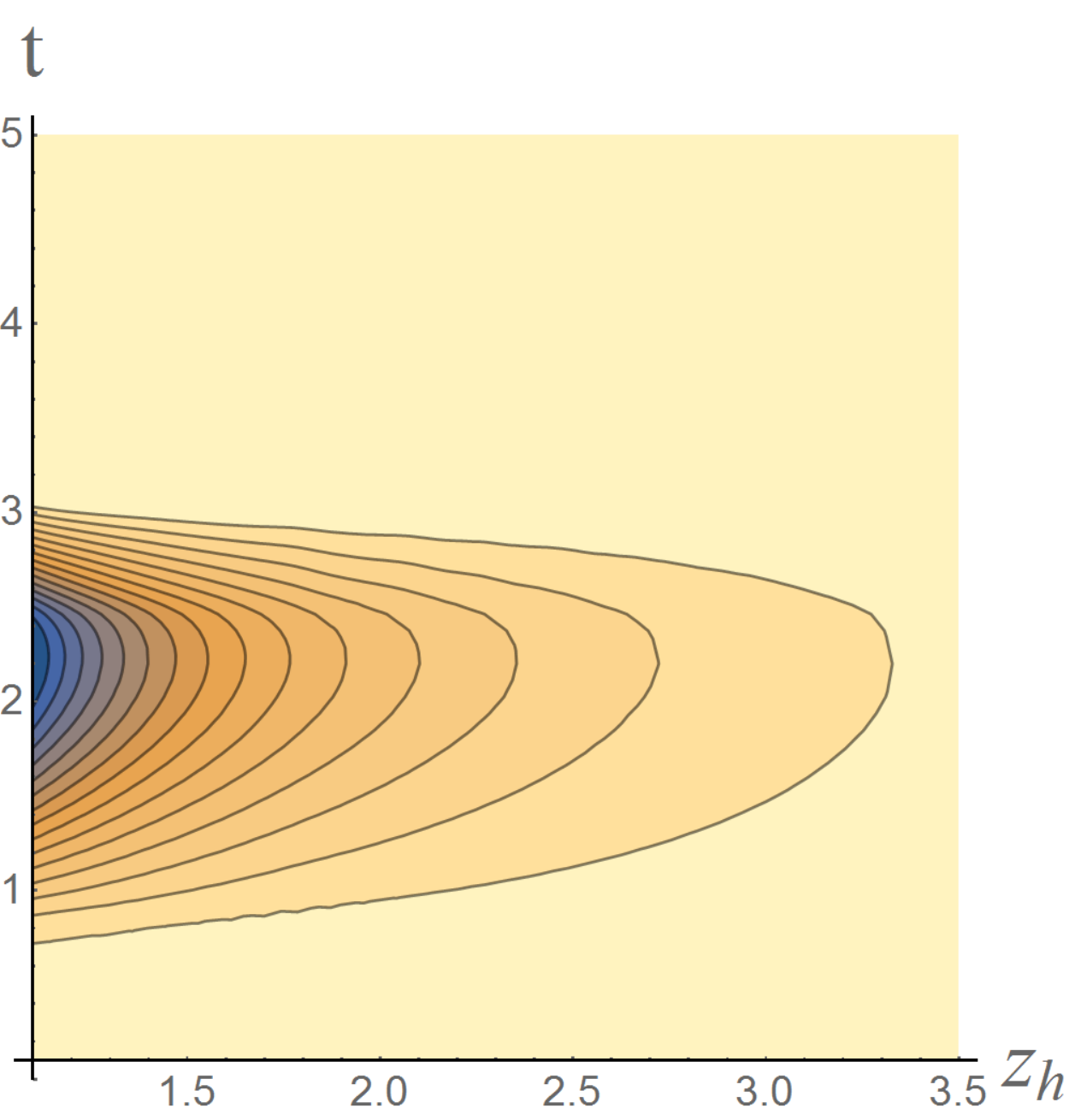}
  \caption{Contour plots of the mutual information for the system of two intervals for the fixed initial temperature at $z_H=4$, while the final one varies. In the left plot: intervals of different lengths
  $\ell_1=4$, $\ell_2=6$ and $x=1.68$; in the right plot intervals of equal length $\ell_1=5$, $\ell_2=5$ and $x=1.68$.
The light yellow domains correspond to the vanishing  mutual information.  }
 \label{fig:mutcont1}
\end{figure}

In Fig.\ref{fig:mutcont1}  the mutual information time evolution  for different  final temperatures  
for equal and different lengths of two intervals are presented.  We see that  the wake up time is smaller and scrambling time is bigger for the equal intervals.

\begin{figure}[h!]

\centering
\begin{picture}(185,125)
\put(-100,0){\includegraphics[scale=0.3]{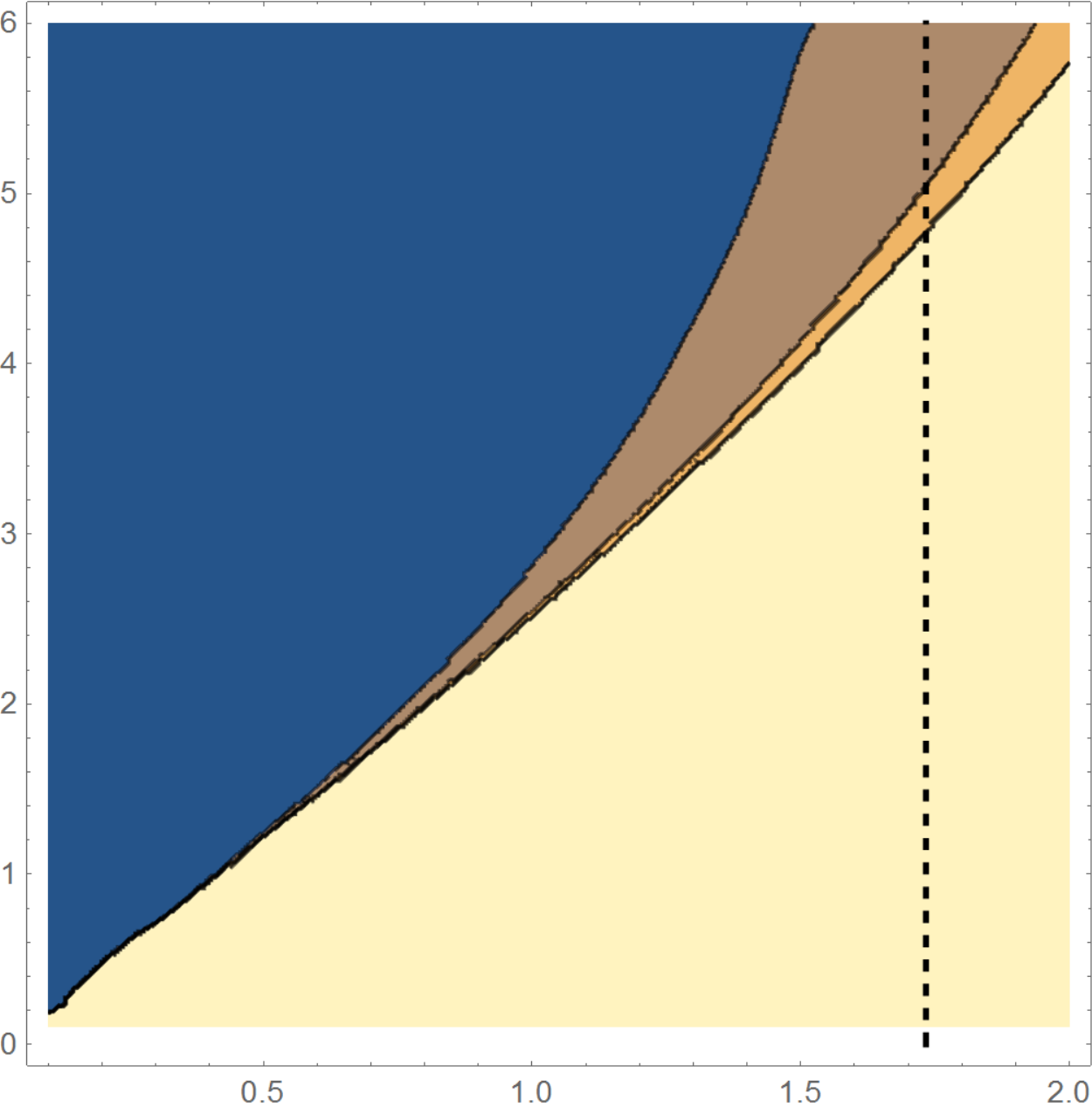}$\,\,\,\,\,\,\,\,$
 \includegraphics[scale=0.3]{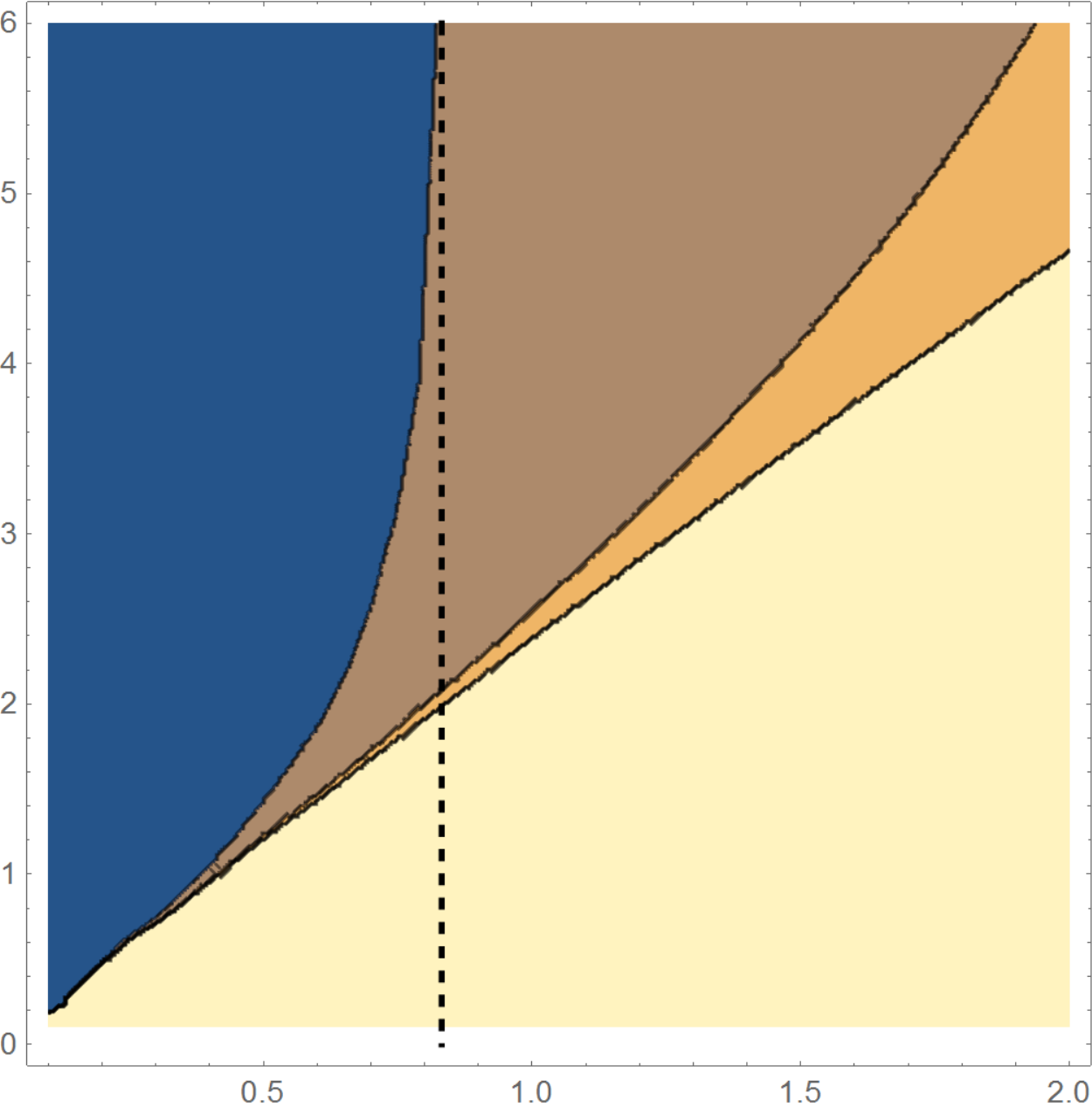}$\,\,\,\,\,\,\,\,$
\includegraphics[scale=0.3]{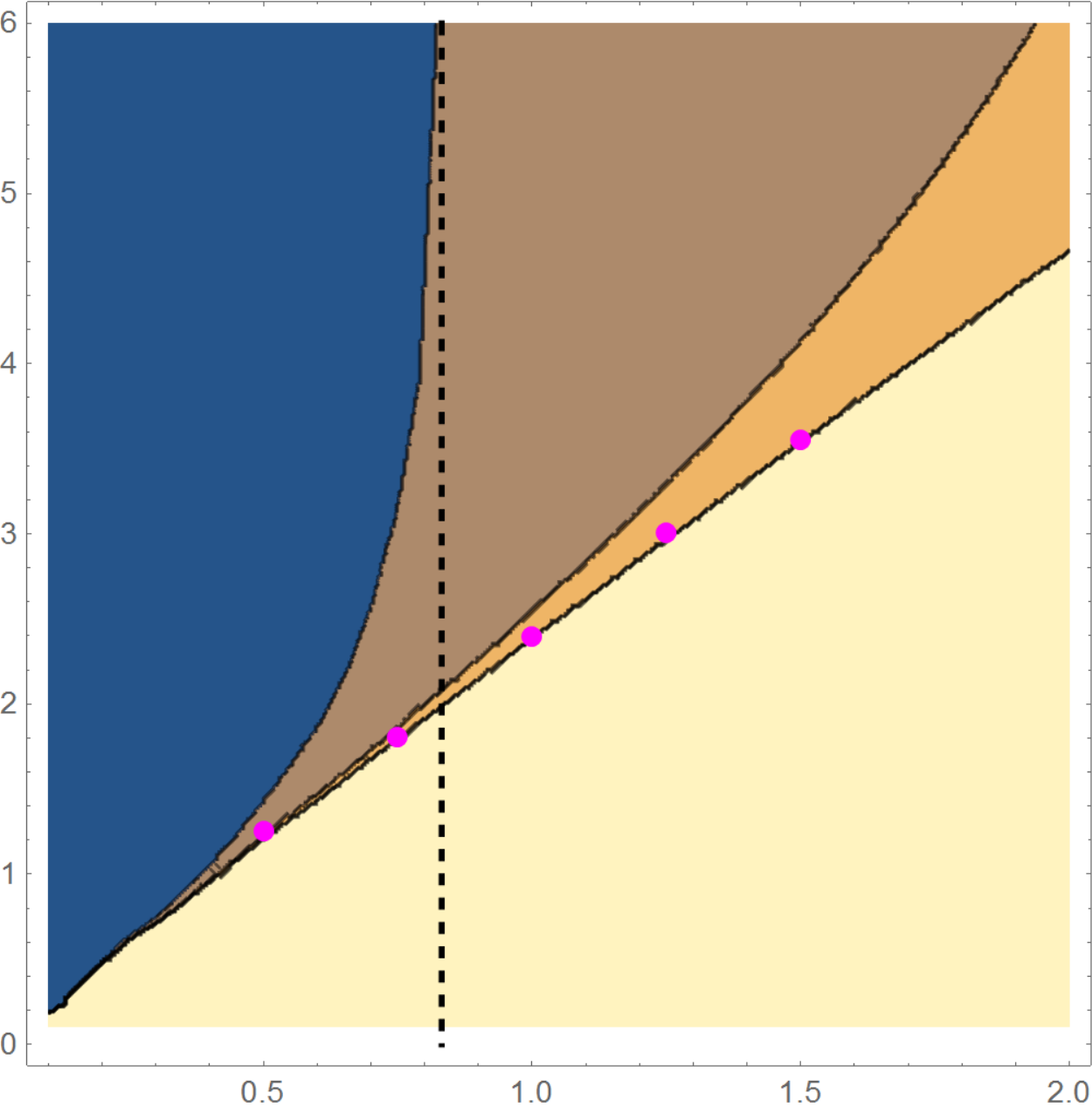}}
  \put(200,-10){$x_{scr}$}
    \put(-20,-10){$x_{scr}$}

    \put(70,-10){$x_{scr}$}
      \put(-80,120){$z_H=4,z_h=2.5$}
       \put(40,120){$z_H=4,z_h=1.2$}
        \put(170,120){$z_H=4,z_h=1$}
 \end{picture}\\
 \begin{picture}(185,150)
\put(-100,0){
  \includegraphics[scale=0.42]{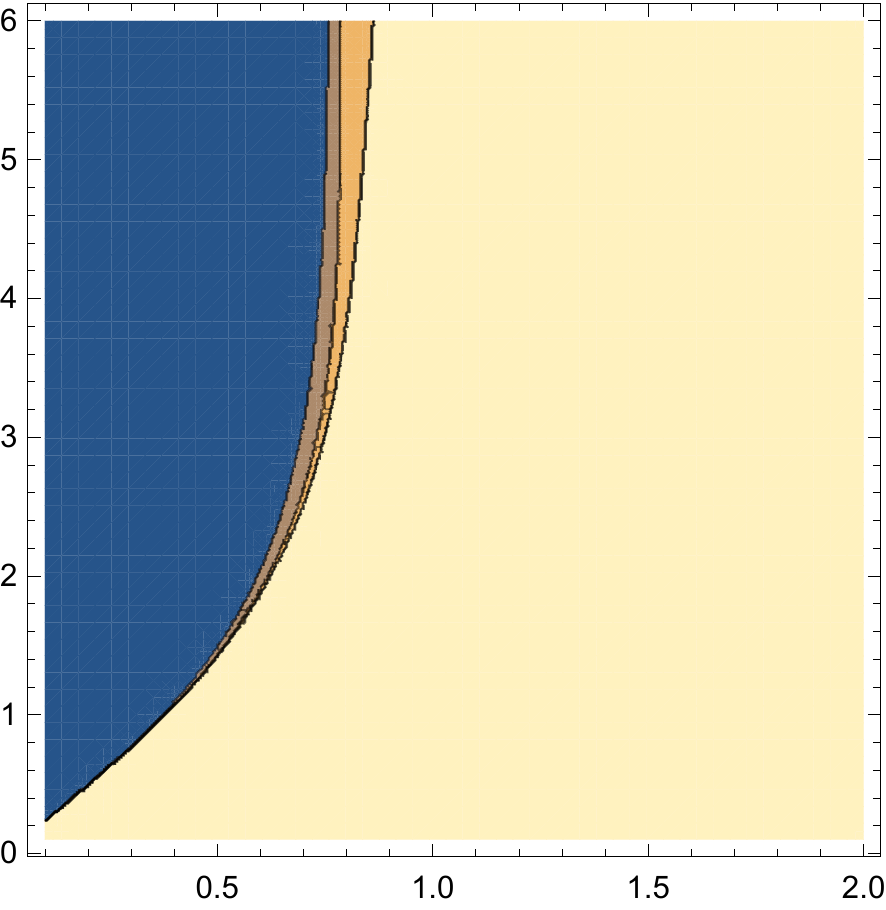}$\,\,\,\,\,\,\,\,$
    \includegraphics[scale=0.3]{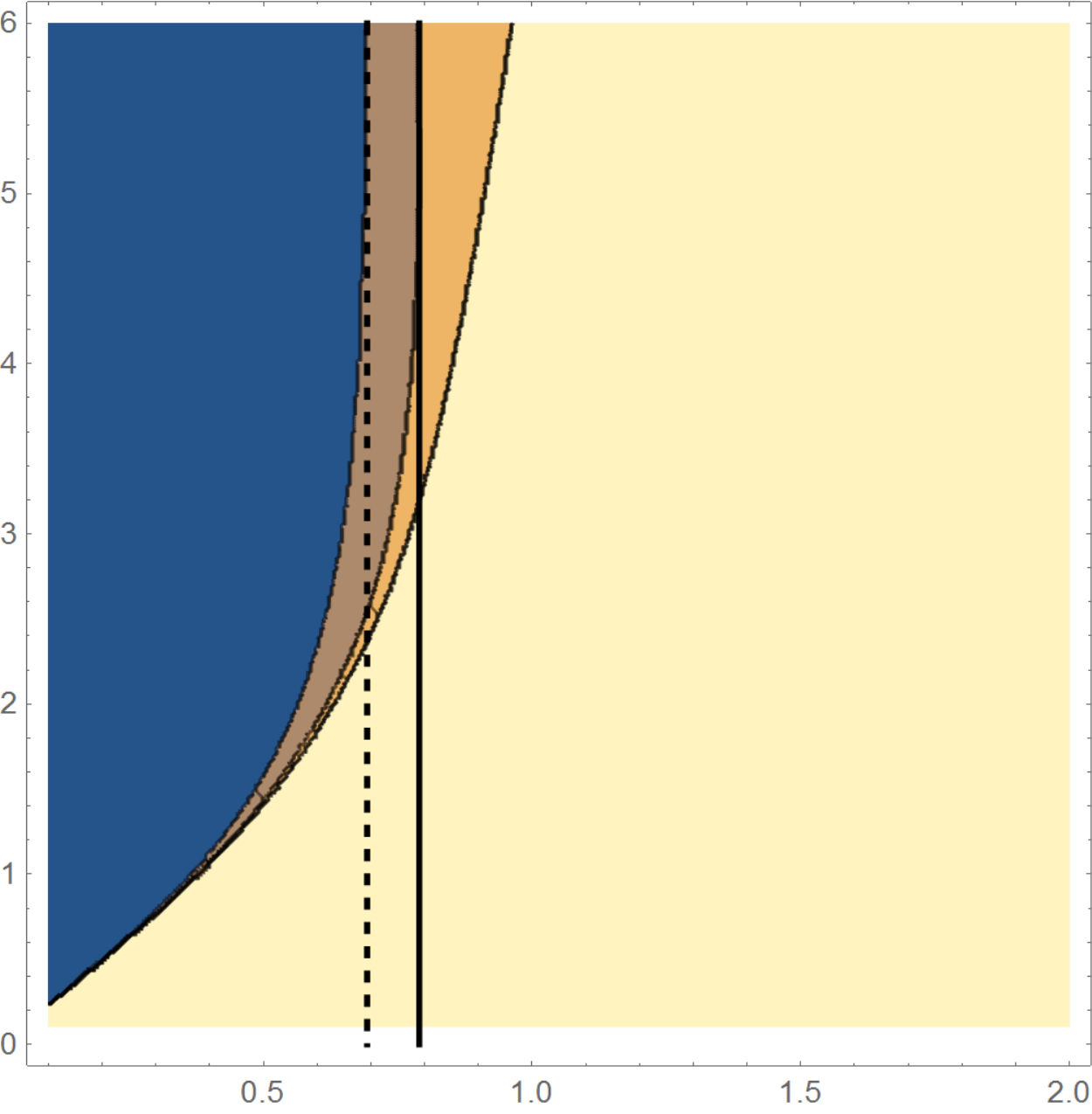}$\,\,\,\,\,\,\,\,$
        \includegraphics[scale=0.3]{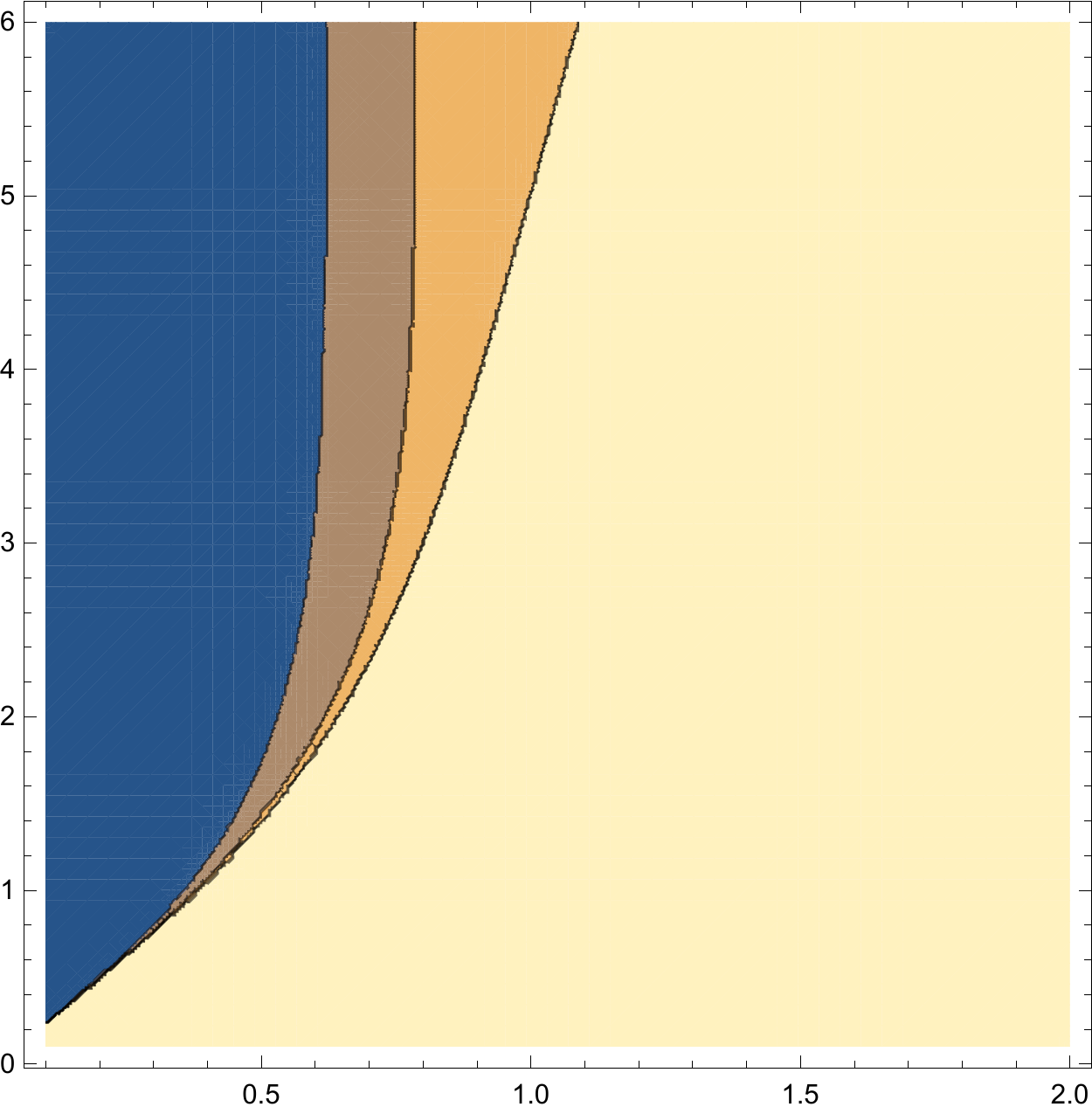}}
\put(50,-10){$x_{scr}$}
 \put(75,-10){$x_{wup}$}
   \put(-80,120){$z_H=1.2,z_h=1.1$}
     \put(40,120){$z_H=1.2,z_h=1$}
        \put(170,120){$z_H=1.2,z_h=0.9$}
            
 \end{picture}
\caption{Zones of different regimes of the mutual information behaviour for different 
$\ell$ and separation $x$. The vertical axes correspond to $\ell$ and the  horizontal axes to $x$. 
The black solid line corresponds to the critical
   value of $x_{scr}=z_h\log 2$, where the scrambling occurs and the dashed one corresponds to  critical value of 
   $x_{wup}=z_h\log 2$.  The different colors correspond to the different regimes presented in Fig.\ref{fig:evol}. In the up plots  $z_H=4$ and   $z_h=2.5, 1.2,1$ (from the left to the right),  in the bottom plots $z_H=1.2$ and $z_h=1.1, 1, 0.9$ (from the left to the right)}
 \label{fig:wakeup}
\end{figure}
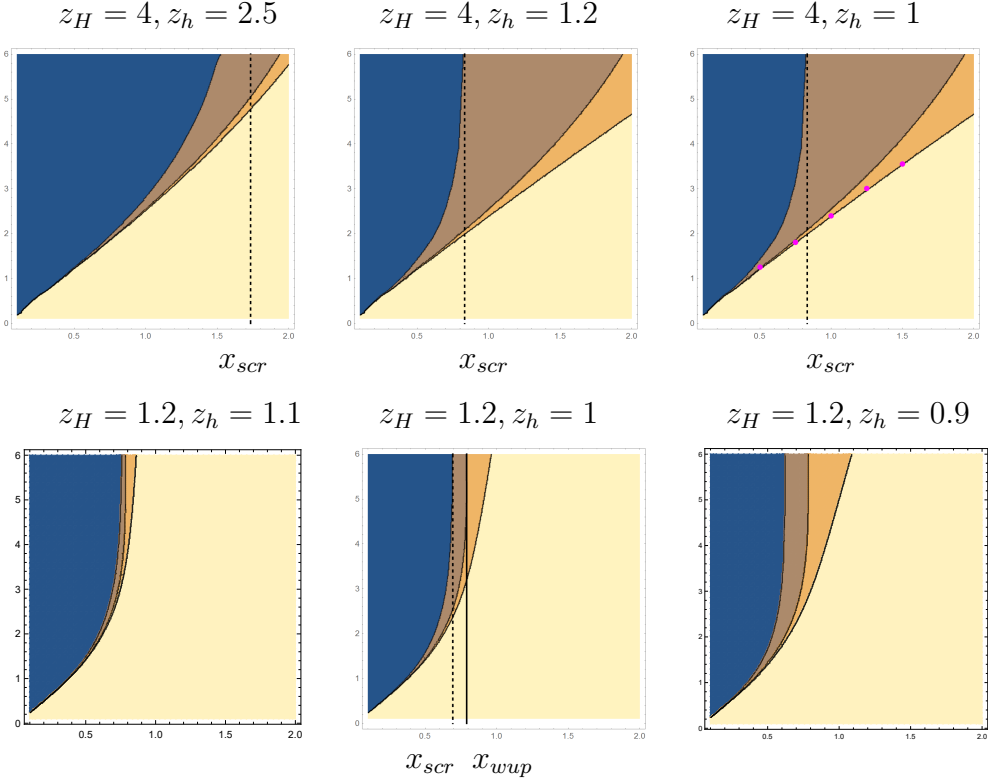

 In what follows we consider a special case $\ell_1=\ell_2=\ell$. In this case our parametric region is   the 2-dimensional one $(x,\ell)$, $\ell>0,x\geq 0$. It can   be divided in
 four zones where one of the four  typical regimes is realized, see Fig.\ref{fig:wakeup}. It is obvious that for large enough  $x$ and fixed $\ell$ the mutual information vanishes, as well as for  large  $\ell$ and small enough $x$ the mutual information behaves as shown by the blue line in Fig.\ref{fig:evol}. The scrambling behavior occurs  at the  parameters for which the  stationary  mutual information of the whole system  vanishes at  the final temperature.  The line in the $(x,\ell)$-plane, where the mutual information at the temperature $T_f$ vanishes, is given by the following equation
\be\label{line}
\ell(x,z_h)=2 z_h \text{arccoth} \left(\sqrt{2}
   \sinh\left(\frac{x}{2 z_h}\right)^{-1}-\coth \left(\frac{x}{2 z_h}\right)\right),
\ee
where $T_f$ is related  with $z_h$ as usual, see  eq.\eqref{temperatures}.

For small $x$ we have 
\be
\ell=(1+\sqrt{2})x+\frac{24+17\sqrt{2}}{24z_h^3}x^3+{\cal O}(x^5).
\ee
and at large $\ell$ this curve approaches from the left to the vertical line 
\bea
x \underset{\ell \to \infty}{\to}x_{scr}&=&z_{h} \ln 2\, ,
\eea 

At $0<t<<t_{heat}$ the temperature of the system is defined by $z_H$ and the line separating 
the region of the bell shape mutual information from the scrambling regime  is given by formula \eqref{line} with $z_H$ instead of $z_h$, and therefore at large $\ell$
\bea
x_{wup}&=&z_{H}\ln{2}\,, \eea 
see Fig.\ref{fig:wakeup}, where the general structure of scrambling and bell regions is presented.

\subsection{Configurations with the bell-type mutual information evolution }

In this section we study in more details the  region of parameters $\ell,x$, where the bell-type of the time evolution of the mutual information during the heating process is realized.
As has been mentioned above, this type of evolution of two disjoint intervals  system (with parameters $\ell,x$) is  such, that the mutual information is zero in the start of the heating process, then it becomes non-zero at the time moment $t_{wup}$,  the so-called wake-up time,  and then it vanishes again at the time moment $t_{scr}$, the scrambling  time. 
For this purpose  we study the scrambling and wake up times for different $\ell$ and $x$.

\begin{figure}[h!]
\centering
\centering\begin{picture}(185,190)
\put(-120,0){ \includegraphics[scale=0.5]{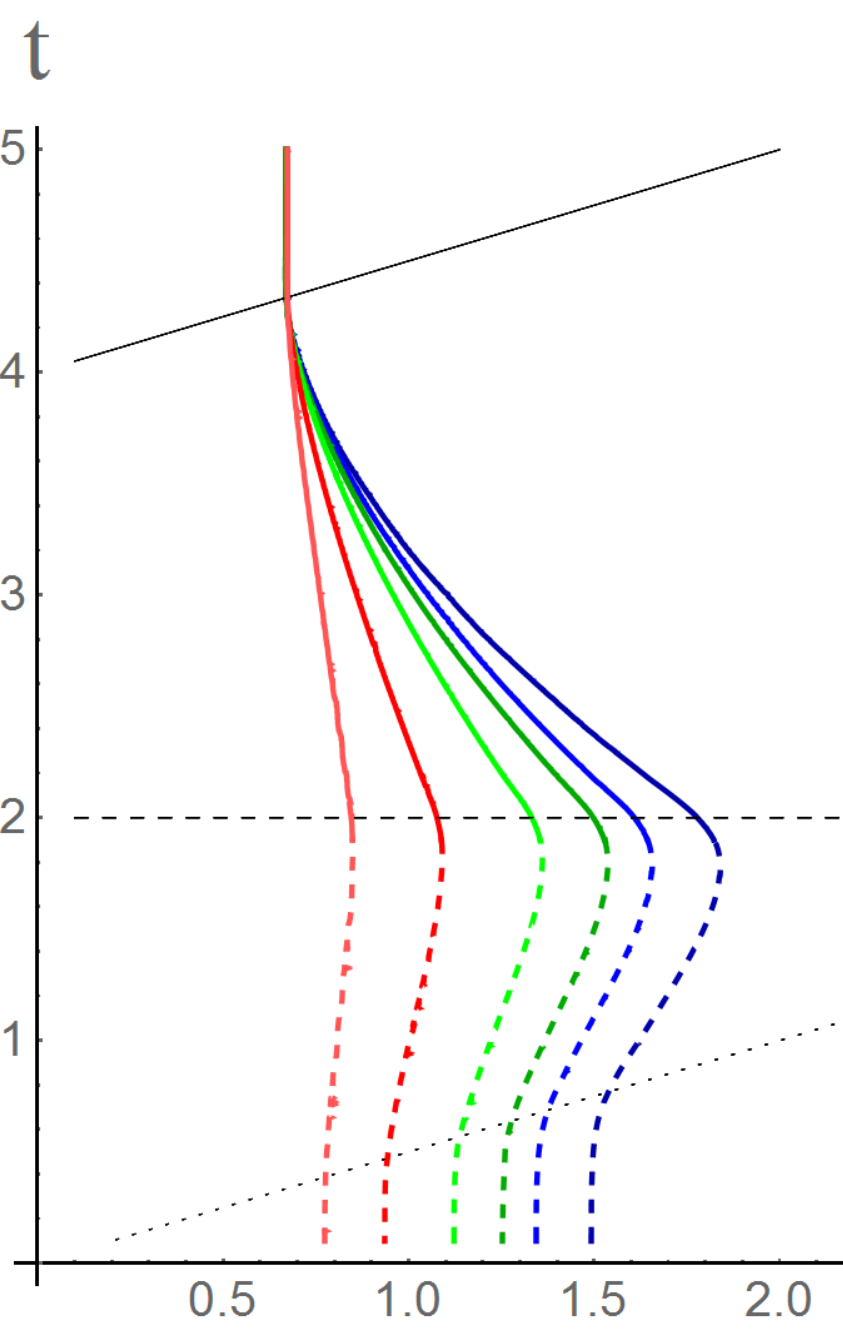}$\,\,\,\,\,\,\,\,$$\,\,\,\,\,\,\,\,$$\,\,\,\,\,\,\,\,$$\,\,\,\,\,\,\,\,$
}
\put(40,0) {\includegraphics[scale=0.45]{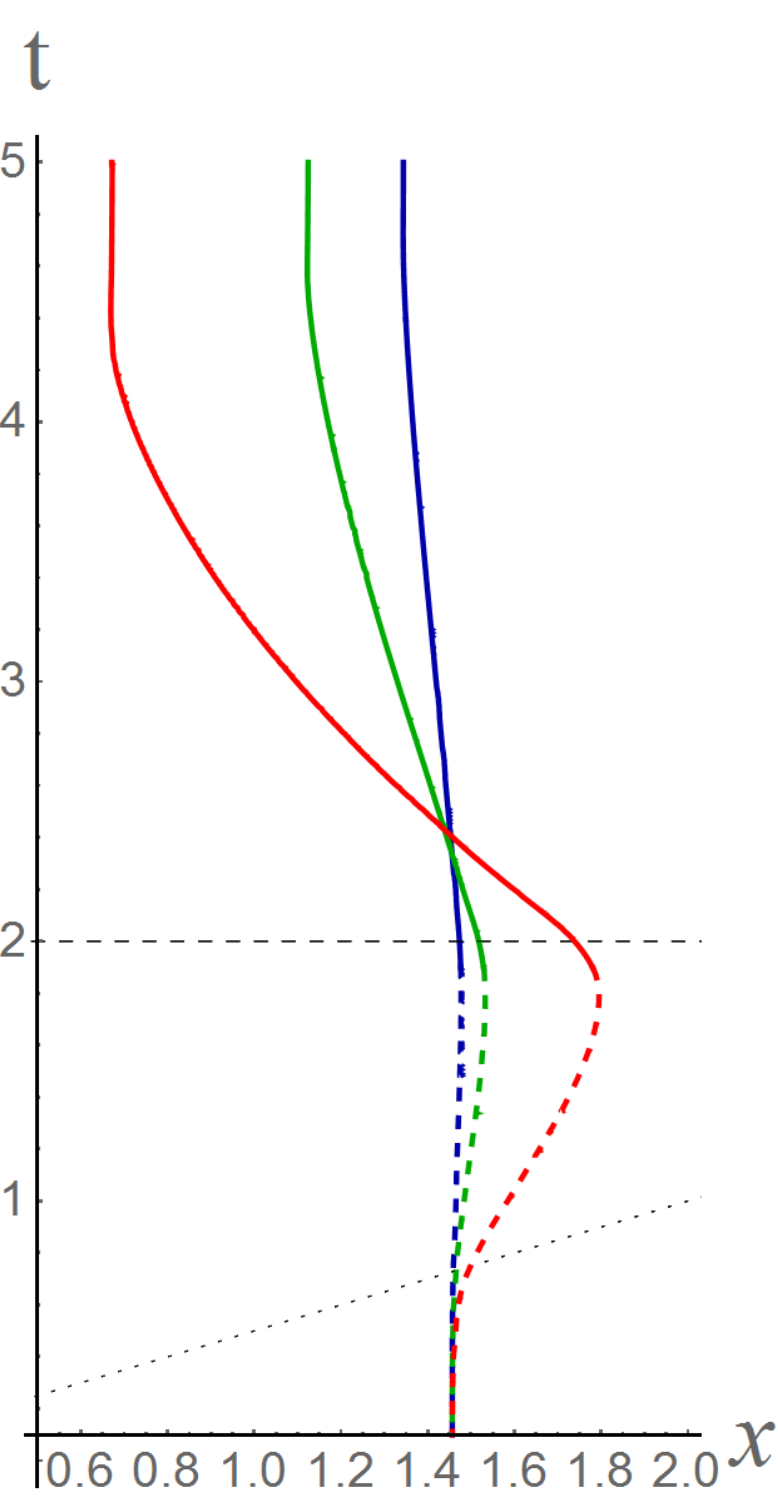}}
\put(170,0) {\includegraphics[scale=0.45]{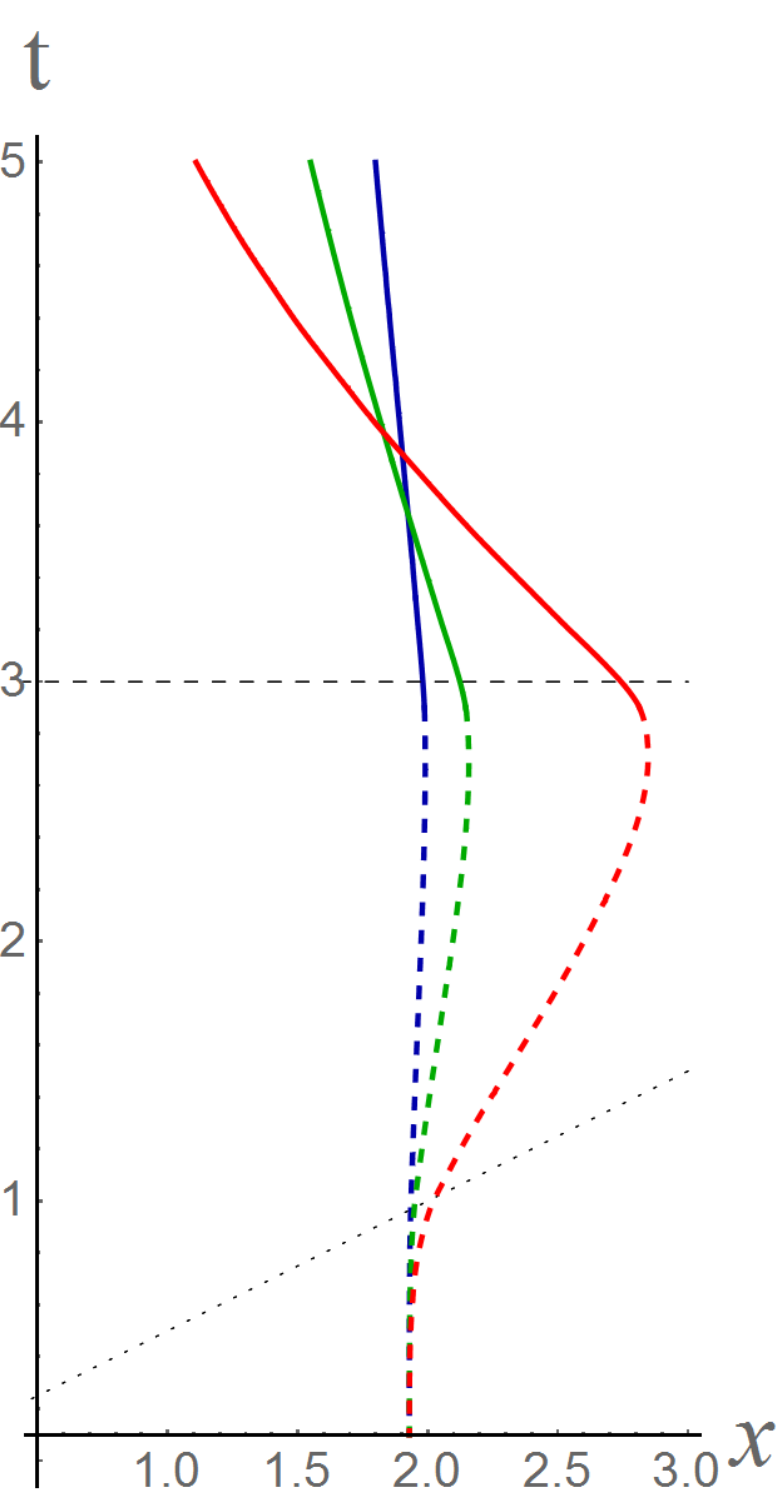}}
    \put(-90,180){$ z_h=1, \ell=4$}  
      \put(70,180){$ z_H=4, \ell=4$}  
       \put(200,180){$ z_H=4, \ell=6$}  
    \put(-61,-10){\vector(1,0){10}}
     \put(-51,-10){\vector(-1,0){10}}
     \put(-51,-10){\line(0,1){85}}
  \put(-60,-10){\line(0,1){25}}
  \put(-21,-10){\line(0,1){85}}
  \put(-38,-40){\line(0,1){52}}
   \put(223,-10){\line(0,1){15}}
 \put(119,-8){\line(0,1){170}}
    \put(98,-10){\line(0,1){25}}
         \put(255,-8){\line(0,1){170}}
     \put(-105,-8){$x_{cr,scr}$}
    \put(-20,-10){\vector(-1,0){18}}
     \put(-30,-10){\vector(1,0){8}}
     \put(-65,-23){$\Delta x_1$}
      \put(-32,-23){$\Delta x_2$}
     \put(-35,-40){$ x_{cr,wup}$}
         \put(-19,15){$ x_0$}
           \put(120,15){$ x_0$}
             \put(255,15){$ x_0$}
          \put(225,-20){$ x_{cr,wup}$}
           \put(95,-20){$ x_{cr,wup}$}
 \end{picture}
 $$\,$$
  \caption{
  The dependence of the scrambling  and wake up times  on the intervals separation $x$. In the left panel  the
  scrambling time (solid lines)  and the wake up times (dashed lines)  are shown for the  different initial temperatures, $z_H=1.2, 1.5, 2, 2.5, 3, 5$ (from the left to the right), and at the fixed final temperature, $z_h=1$. In the central and right panels the scrambling time (solid lines)  and the wake up times (dashed lines) are shown for the different final temperatures corresponding to $z_h=1,2$ and $3$ (from the left to the right), while $z_H=4$. The intervals size is taken to be $\ell=4$ for left and the central panels and $\ell=6$ for the right panel.  The dotted,  dashed  and solid black   lines show the  thermalization 
time of the intervals of the length $x$, $\ell$ and  $x+\ell$, respectively.
}
 \label{fig:tscrx1}
\end{figure}
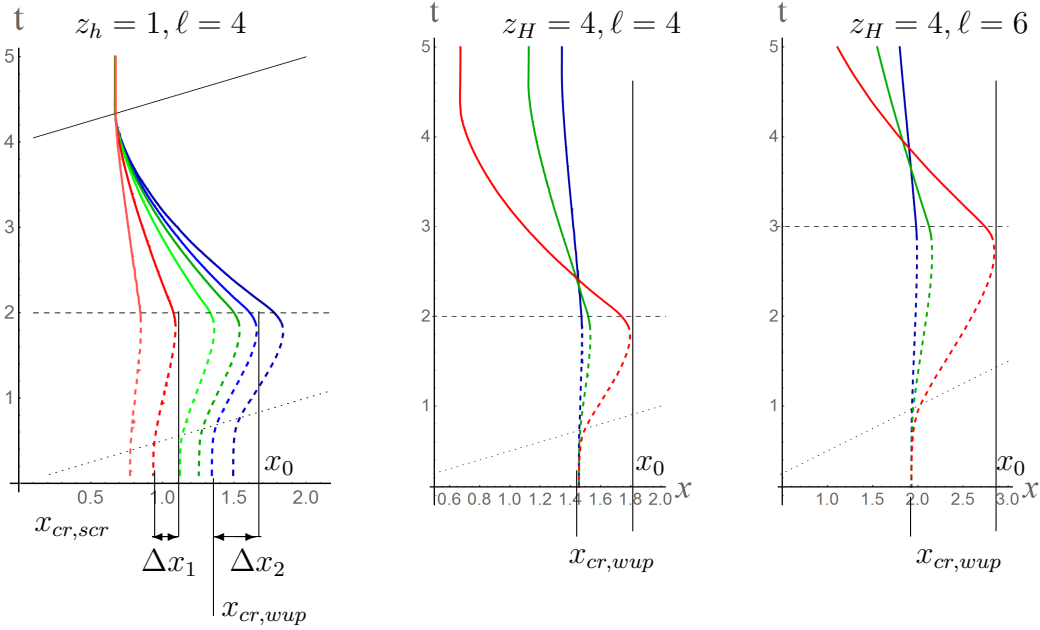

 In Fig.\ref{fig:tscrx1} we plot the dependence of the wake up  times (the dashed lines) and the scrambling times (the solid lines)  on the  intervals separation $x$ for fixed lengths of the  intervals  for different initial
 and  final temperatures. In the left plot of Fig.\ref{fig:tscrx1} the final temperature is fixed and  we see that there is a critical $x_{cr,wup}(z_{H})$ at which the wake up lines start at $t=0$. The value of $x_{cr,wup}(z_H)$ does not depend on $z_h$ and along the  wake up lines $x_{cr,wup}(z_{H})<x_{wup}(t)<x_0(z_{H},z_h)$  is satisfied. The wake up line  disappears at $x_{0}(z_h,z_H)$,   and the scrambling line starts at $x_0$. The scrambling lines satisfy 
 $x_{cr,scr}(z_h)<x_{scr}(t)<x_0(z_{H},z_h)$. We see that  for larger $z_H$ we get a larger $t_{scr}$
 and smaller  $t_{wup}$.  However, we cannot increase $z_H$ too much to get the less $t_{wup}$, since for fixed $x$ there is a critical $x_{cr,wup}(z_{H})$. 

\begin{figure}[h!]
\centering
\begin{picture}(185,190)
\put(-120,0){\includegraphics[scale=0.3]{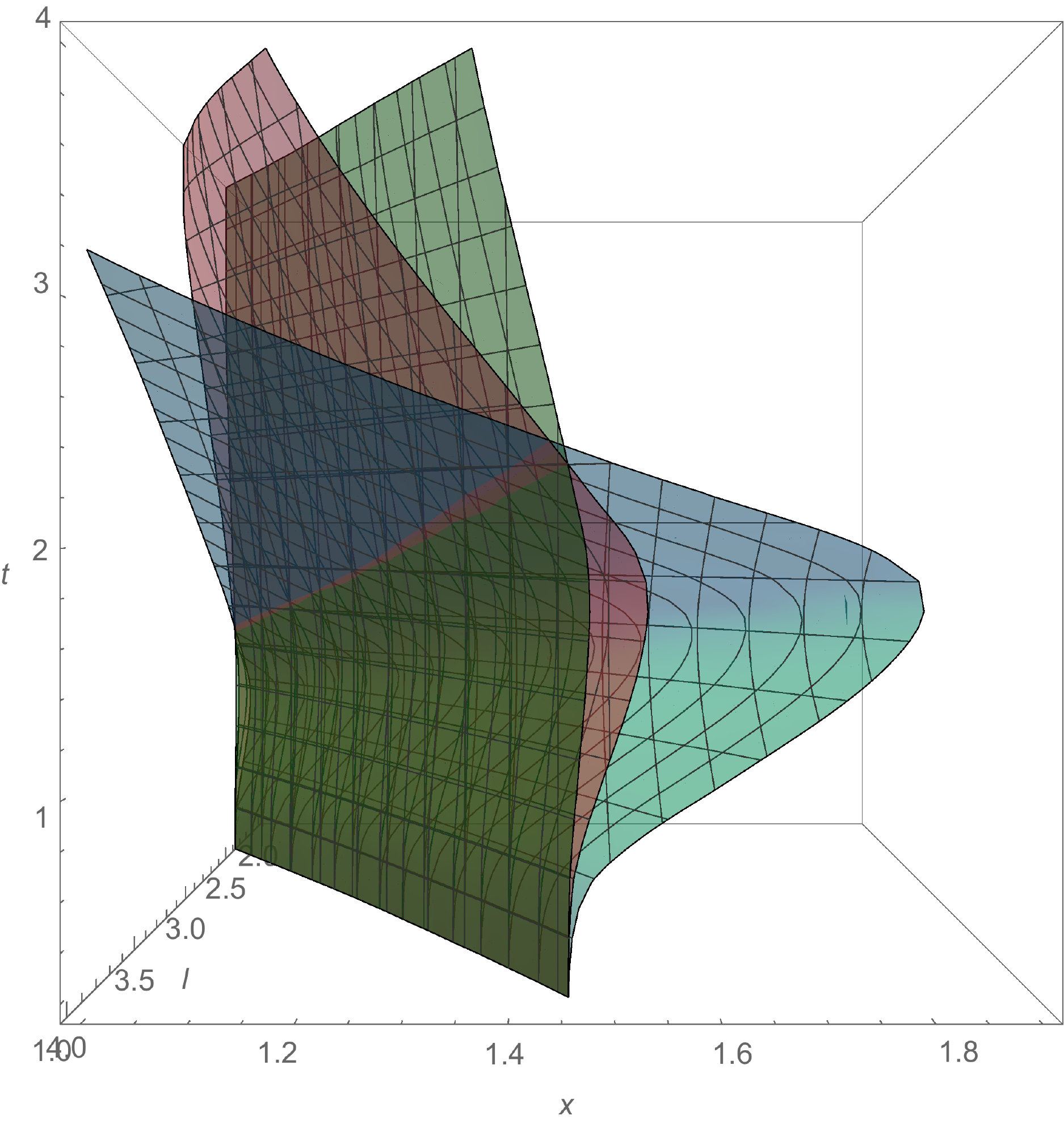}}
  \put(-15,-10){$x$} 
  \put(-120,160){$t$} \put(30,30){$\ell$}\put(65,0){A}
\put(0,25){\line(0,1){80}}
\put(-5,20){\line(1,1){10}}
\put(5,20){\line(-1,1){10}}
\put(0,58){\circle*{2}}
\put(0,83){\circle*{2}}
 \put(5,55){$t_{wup}$}
  \put(5,95){$t_{scr}$}
 \put(305,0){B}
\put(120,5){\includegraphics[scale=0.31]{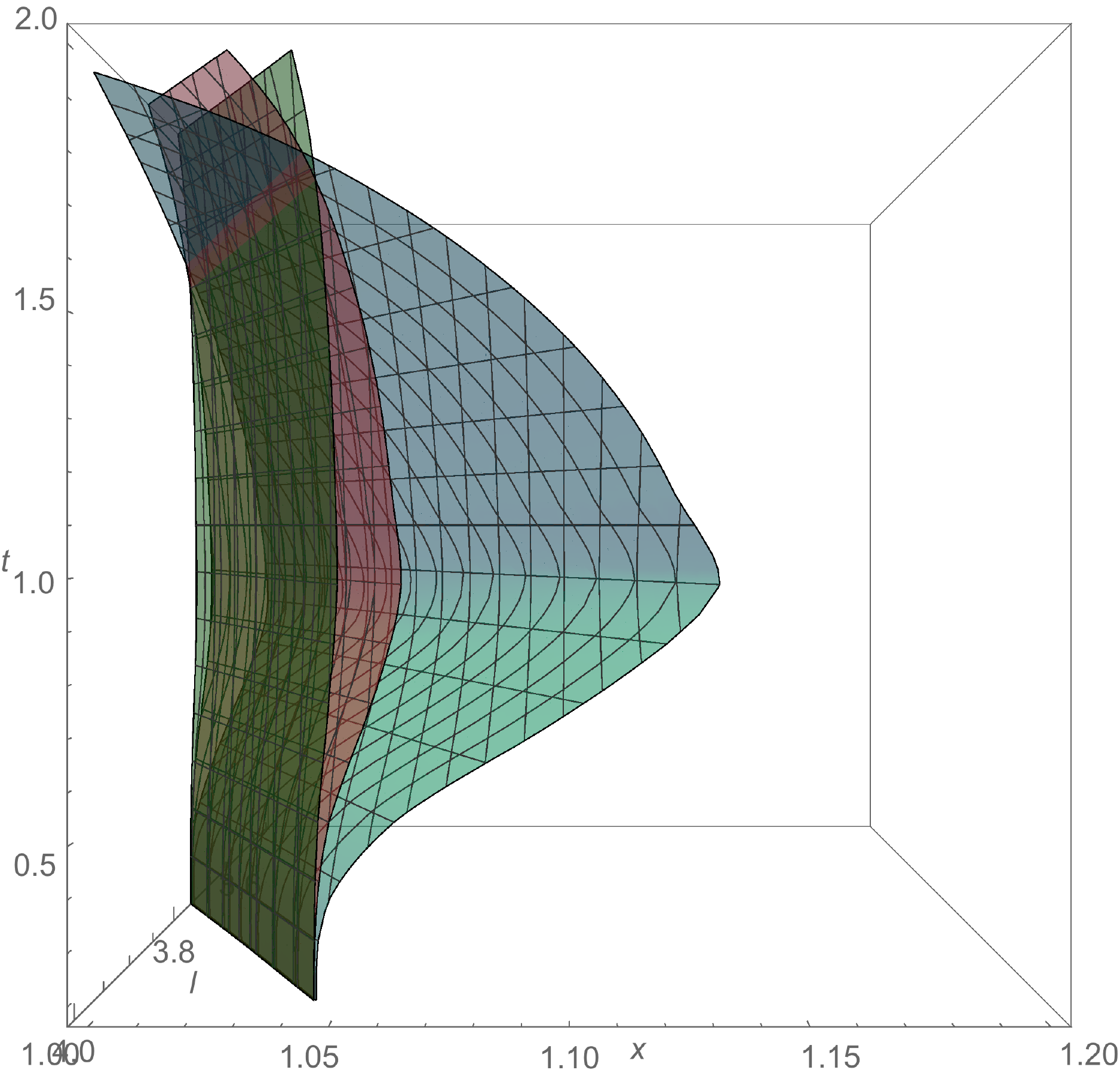}}
 \end{picture}
 \caption{
 The dependence of the  scrambling  and wake up times for the fixed initial temperature ($z_H=4$) on $x$,  and A) equal lenghs $\ell_1=\ell_2=\ell$  and B)  for non-equal lenghts $\ell_1=2\ell_2$.
 The final temperatures correspond to  $z_h=3,2,1$ (green, pink and cyan surfaces, respectively) }
 \label{fig:3D-wup-scr}
\end{figure}
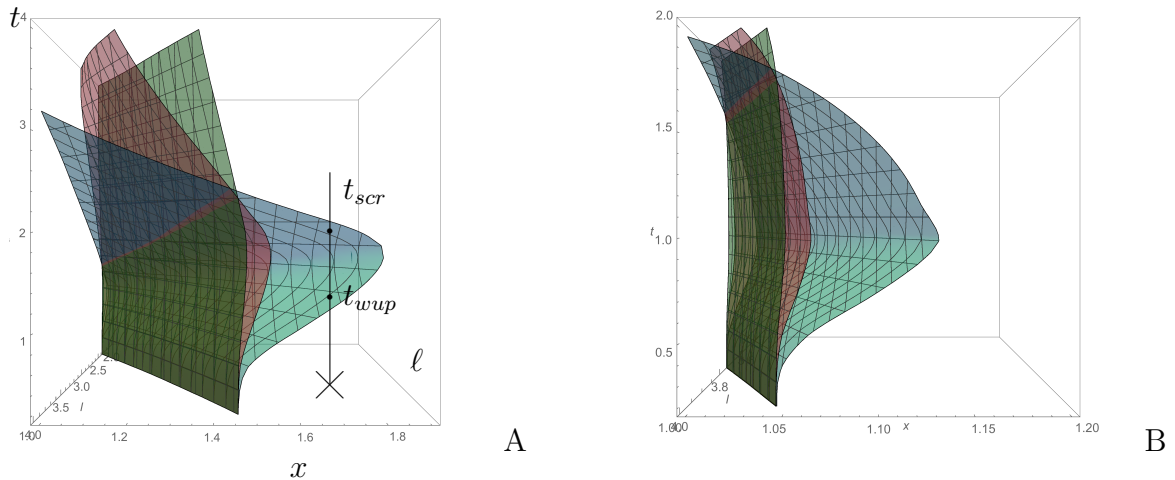

 We see, that for a small  difference between the final and initial temperatures
 only in a very narrow interval $\Delta x$   we can get  the scrambling and wake up time simultaneously, i.e. the distance  between two vertical lines in the left plot in Fig.\ref{fig:tscrx1}  decreases when the temperature difference is decreasing. For example, in Fig.\ref{fig:tscrx1}  $\Delta x_1<\Delta x_2$.   Or in other words, a larger signal, i.e. a higher temperature difference, permits  the information exchange for wider range set  of parameters. We also see at the left plot of  
 Fig.\ref{fig:tscrx1}, that $x_{cr,scr}$
 depends only on the $z_h$ and does not depends on the $z_H$. From the right and central  plots of  Fig.\ref{fig:tscrx1} 
 we see that $x_{cr,wup}$ depends  on the initial temperature and $\ell$. Increasing $\ell$ we increase $x_{cr,wup}$,    $x_0$ and $x_0-x_{cr,wup}$.

In Fig.\ref{fig:3D-wup-scr} we present the 3D picture of the scrambling and wake up times for the fixed  initial and final 
temperatures and  varying $\ell_1$, $\ell_2$ and $x$. To find the  scrambling and wake up times corresponding to the given
 $\ell_1$, $\ell_2$ and $x$ we take a vertical line  with fixed $x$ and $\ell$, and find its cross-sections with the surface. If there are two cross-sections then there are $t_{wup}$  and  $t_{scr}$, as shown at Fig.\ref{fig:3D-wup-scr}. One  
 cross-section corresponds to   $t_{scr}$, and their abscense means that we are in the blue zone on Fig.\ref{fig:mutcont1}.  Comparing the left and right plots in 
Fig.\ref{fig:3D-wup-scr} we see that the bell zone is wide for equal intervals as compared to non equal ones, the same behavior on can see in Fig.\ref{fig:mutcont1}.
Taking the different sections $x=const$ of the surface depicted on Fig.\ref{fig:3D-wup-scr} we get  dependences of
the scrambling and wake up times on $\ell$.

 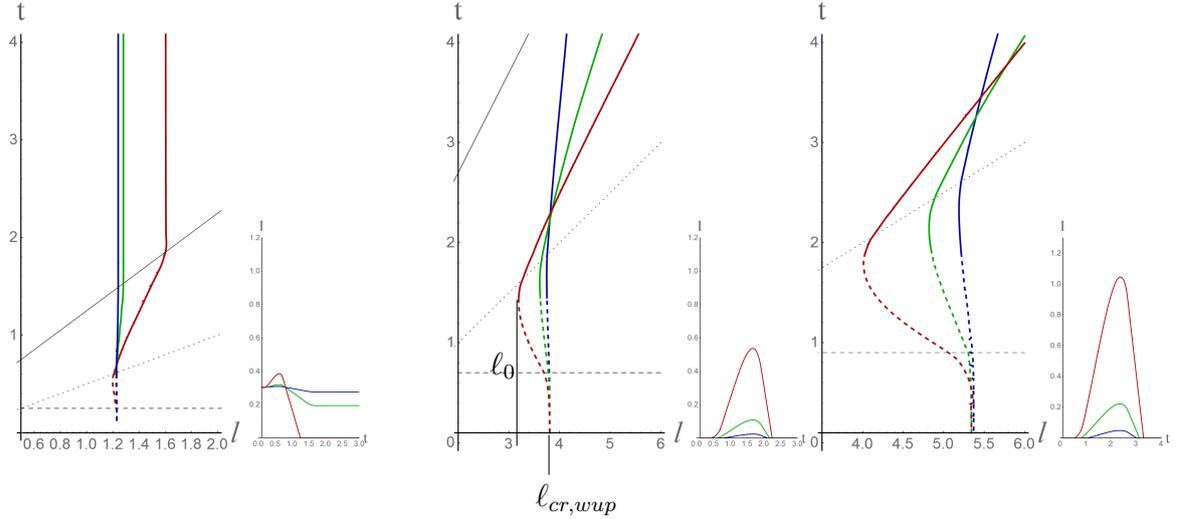
\begin{figure}[h!]
\centering
\begin{picture}(185,150)
\put(-120,0){\includegraphics[scale=0.40]{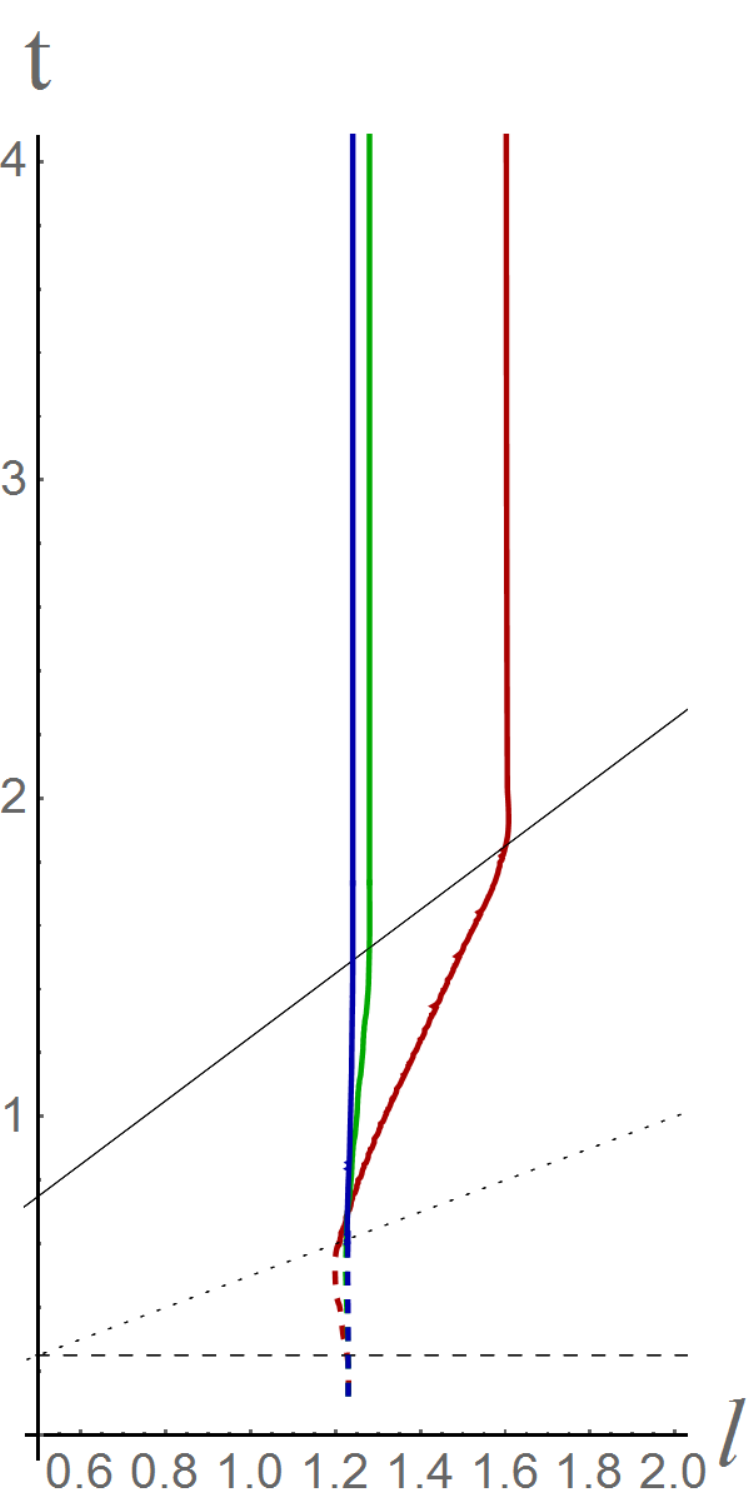}}
\put(-30,2){\includegraphics[scale=0.2]{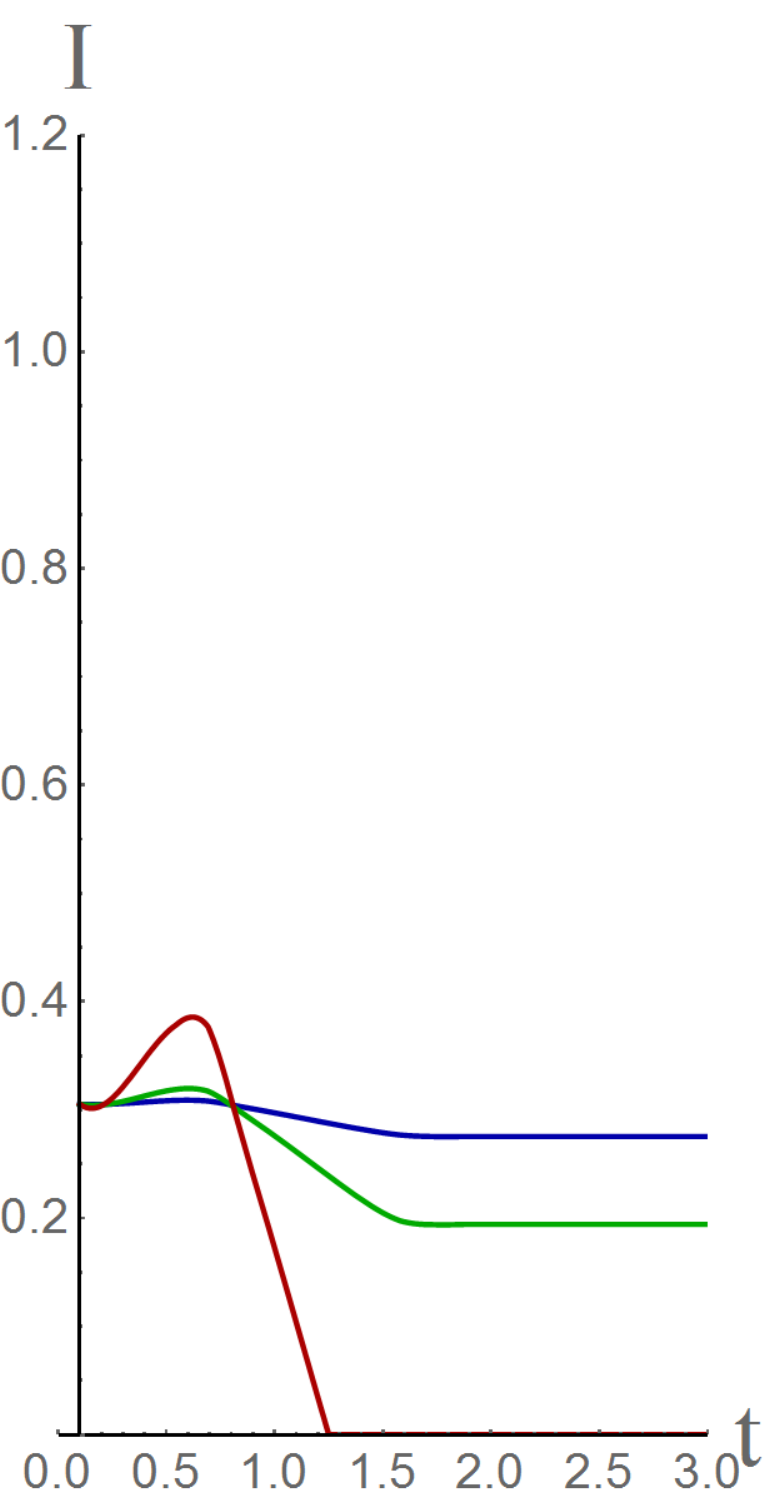}}
\put(40,0){ \includegraphics[scale=0.4]{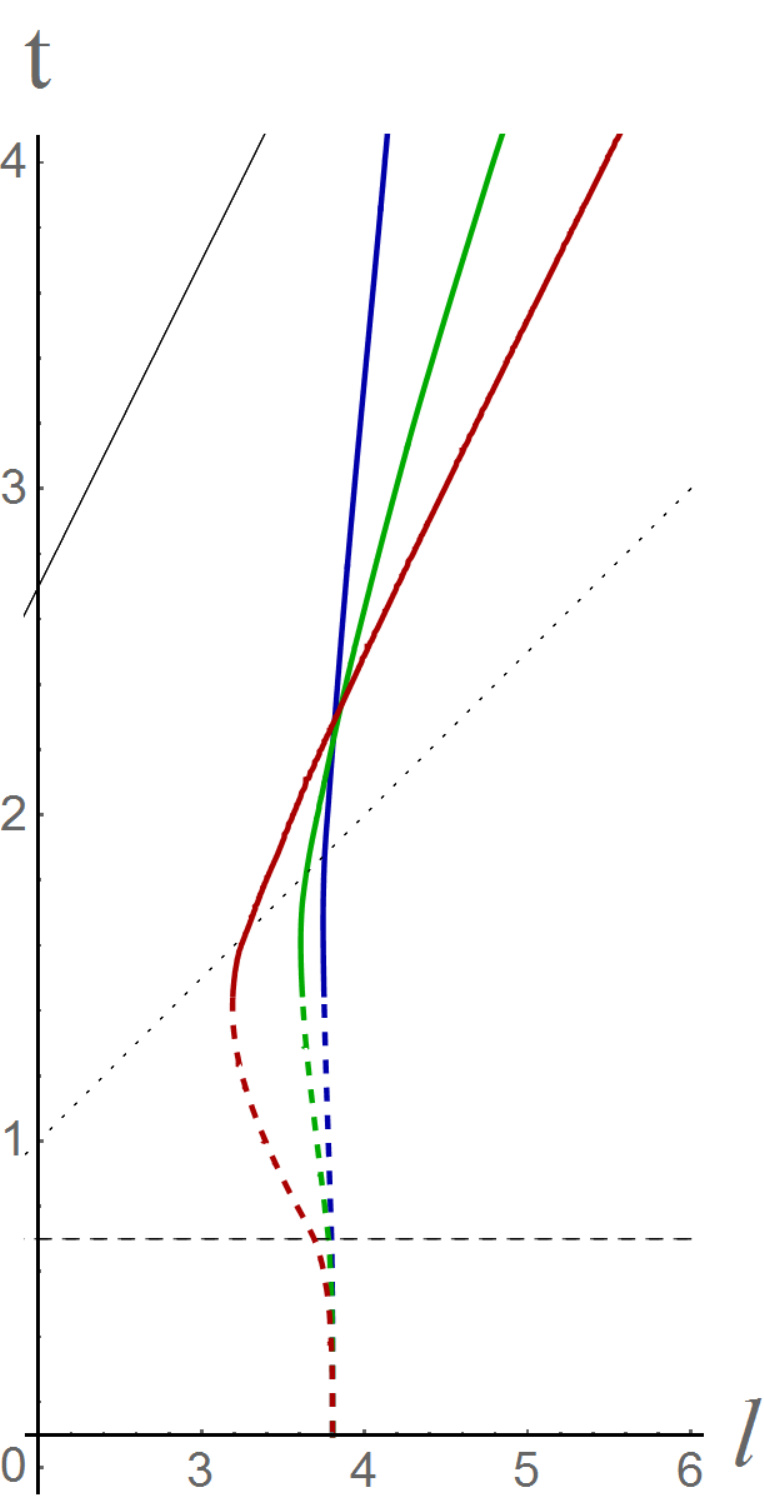}}
\put(134,2){\includegraphics[scale=0.2]{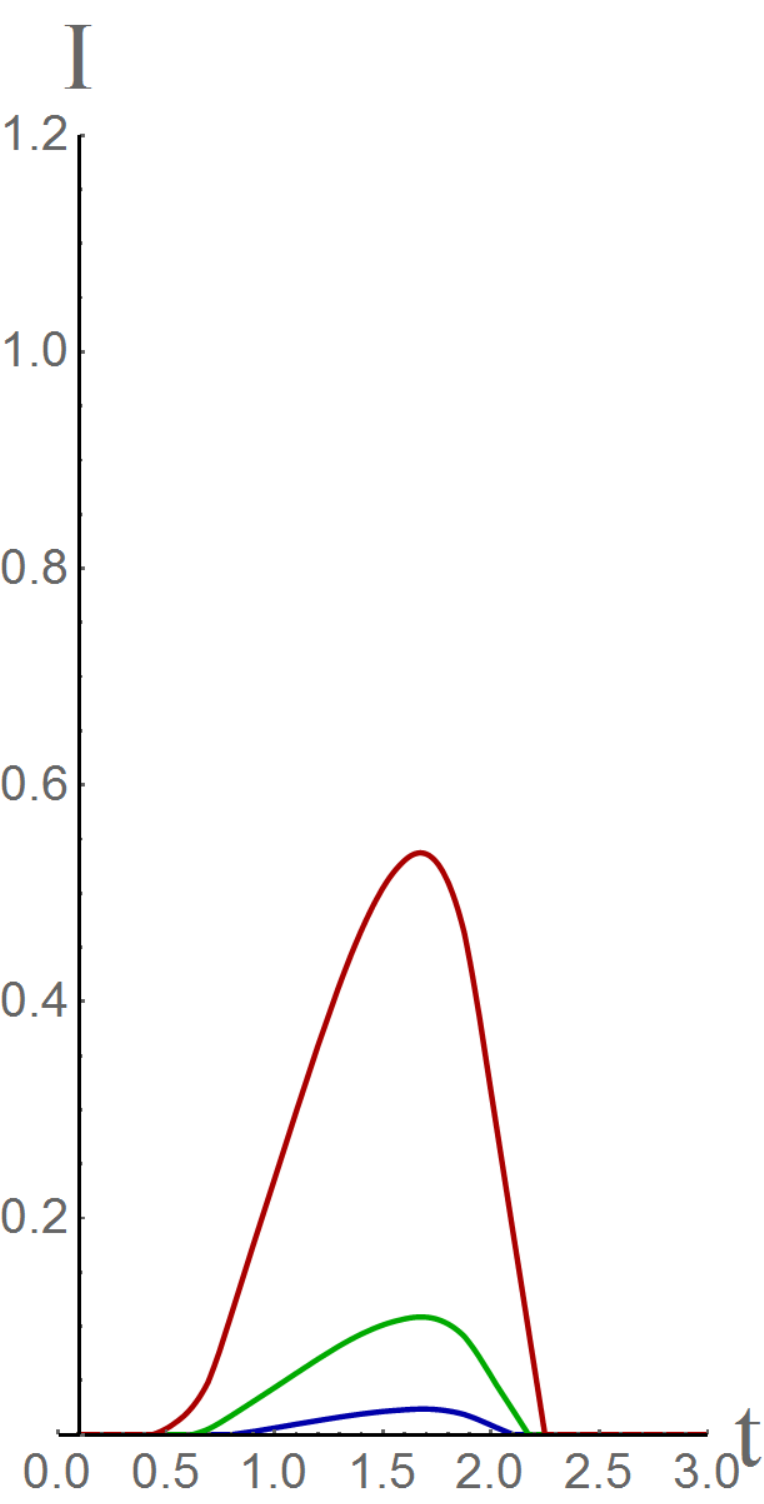}}
\put(180,0){\includegraphics[scale=0.4]{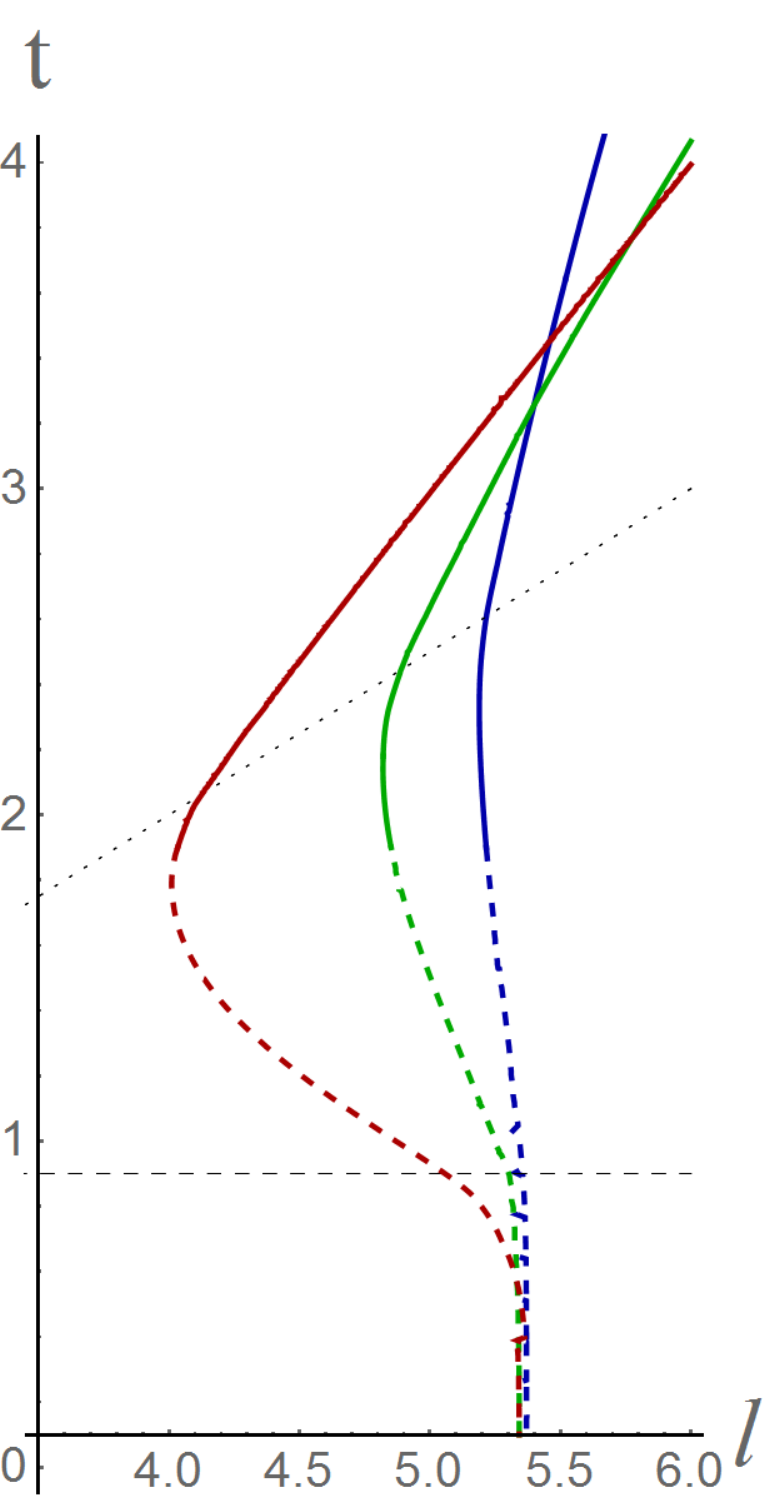}}
\put(270,2){\includegraphics[scale=0.2]{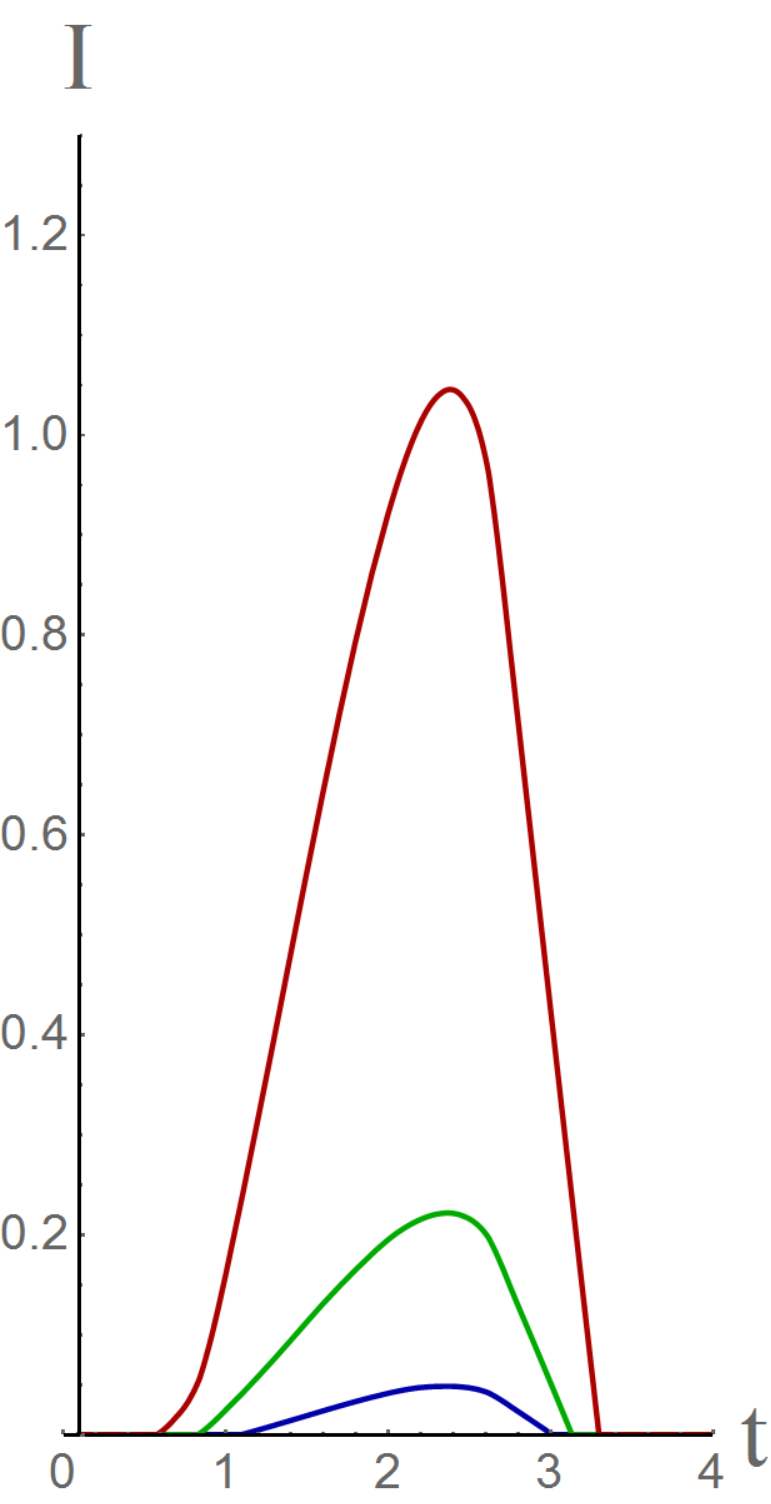}}
 \put(60,30){$ \ell_0$}
          \put(77,-20){$ \ell_{cr,wup}$}
           \put(82,-8){\line(0,1){15}}
         \put(70,3){\line(0,1){55}}
          \end{picture}\\$\,$
\caption{The dependence of the scrambling (solid lines)  and wake up (dashed lines) times  on $\ell$ for the fixed  initial temperature, $z_H=4$ and the final one $z_h$ varying: $z_h=1$ (the red curve), $z_h=2$ (the green one), $z_h=3$ (the blue one). In the left panel $x=0.5,\ell=1.4$, in the central panel  $x=1.4,\ell=3.8$ and in the right one $x=1.8,\ell=5.3$. The mutual information time evolution is presented by the correspondently colored lines
in the right parts of the panels.
}
 \label{fig:tscrl2}
\end{figure}

In Fig.\ref{fig:tscrl2} we plot the  dependence of the scrambling and wake times on the interval size, $\ell_1=\ell_2=\ell$,  for different final temperatures  with the initial temperature being fixed, $z_H=4$.  We see that the wake up time exists  for $\ell_0<\ell<\ell_{cr,scr}$ and that increasing the  distance between intervals $x$ we increase $\ell_0,\ell_{cr,scr}$ and $\ell_{cr,scr}-\ell_0$.

\begin{figure}[h!]
\centering\begin{picture}(185,150)
\put(-120,0){
 \includegraphics[scale=0.35]{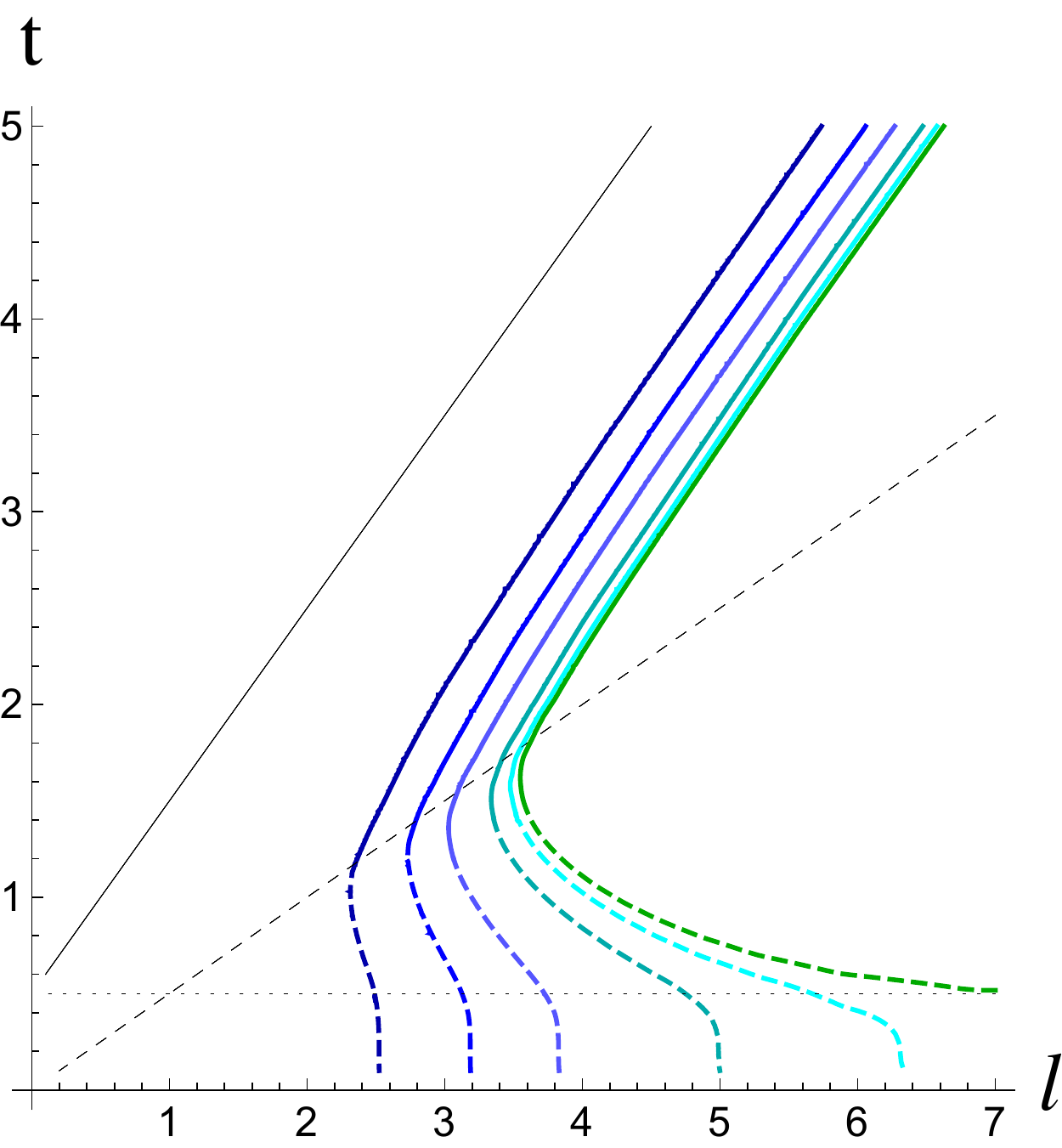} 
 \includegraphics[scale=0.2]{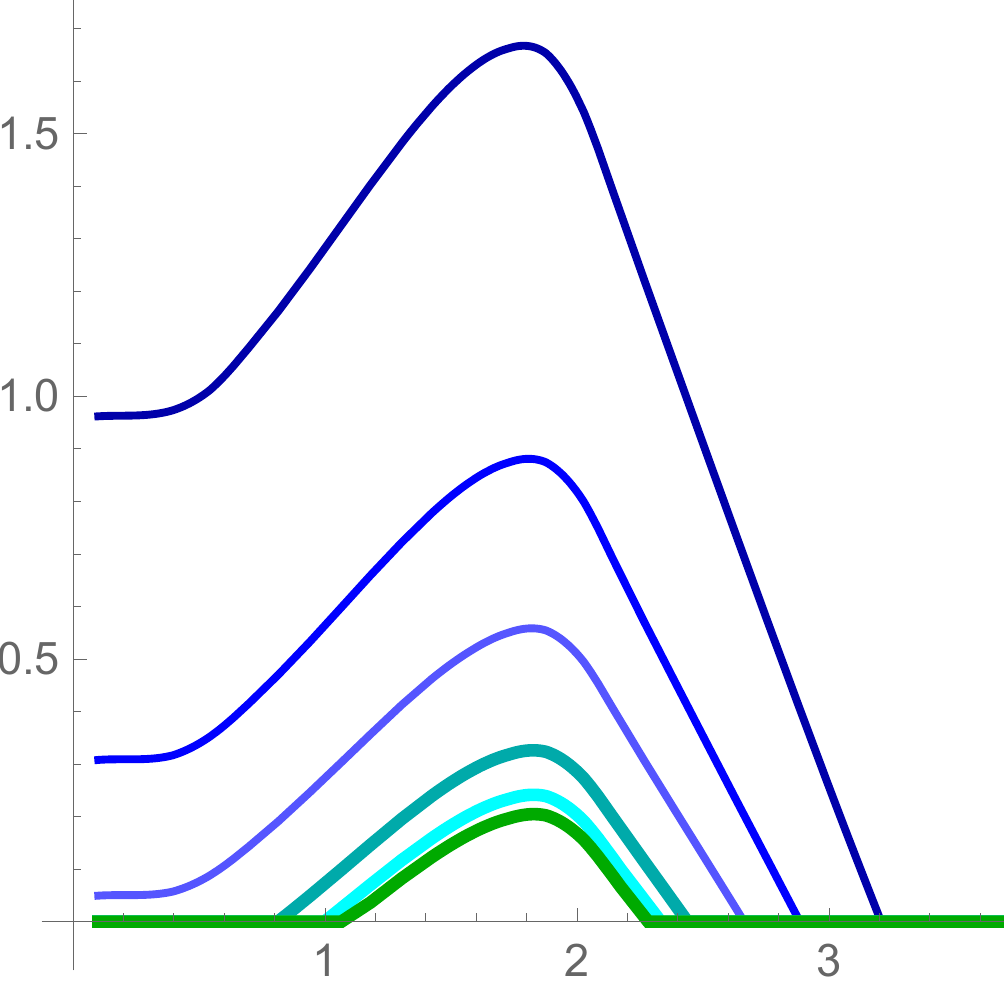} $\,\,\,\,\,\,$$\,\,\,\,\,\,$
 \includegraphics[scale=0.35]{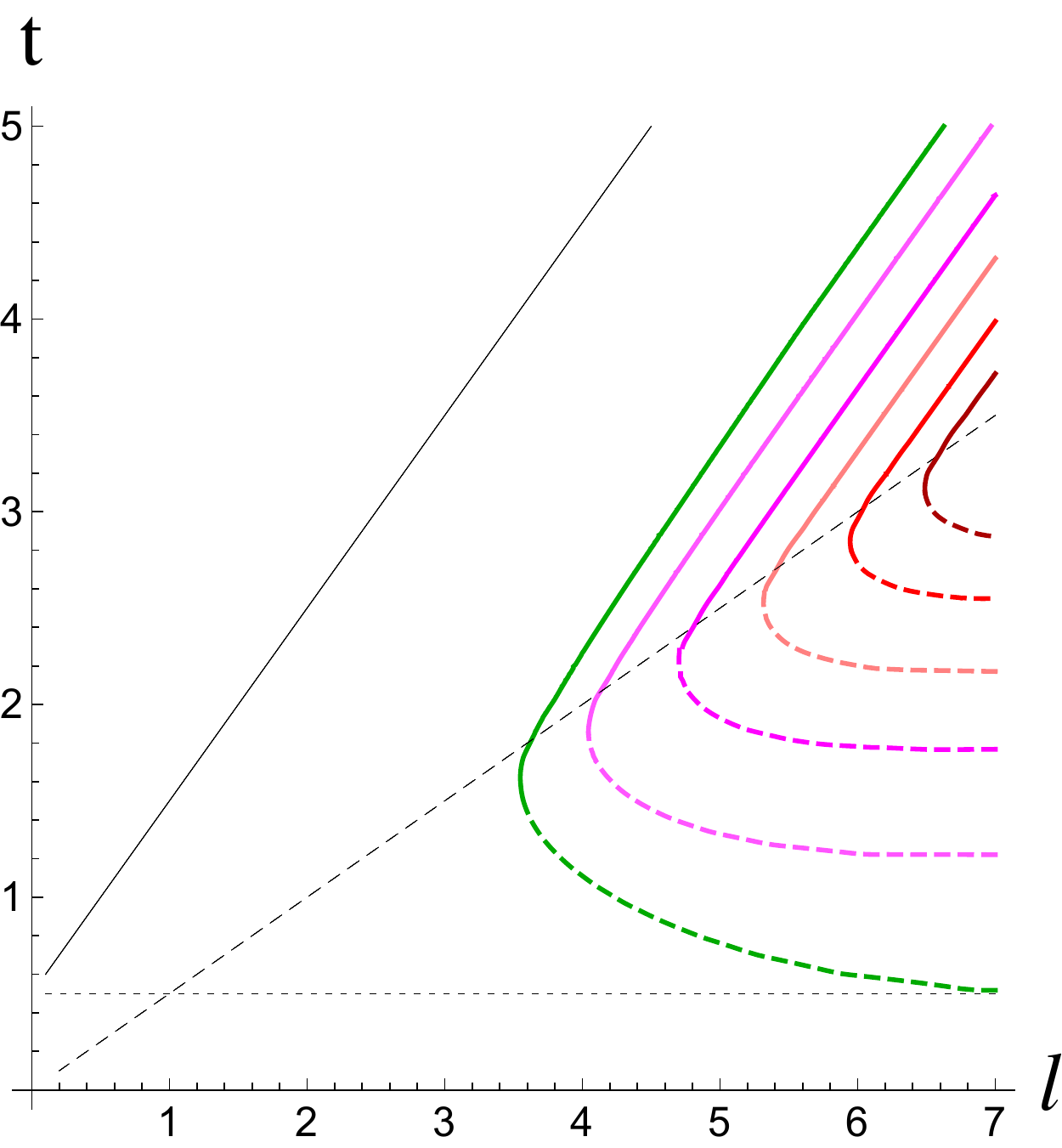} 
 \includegraphics[scale=0.2]{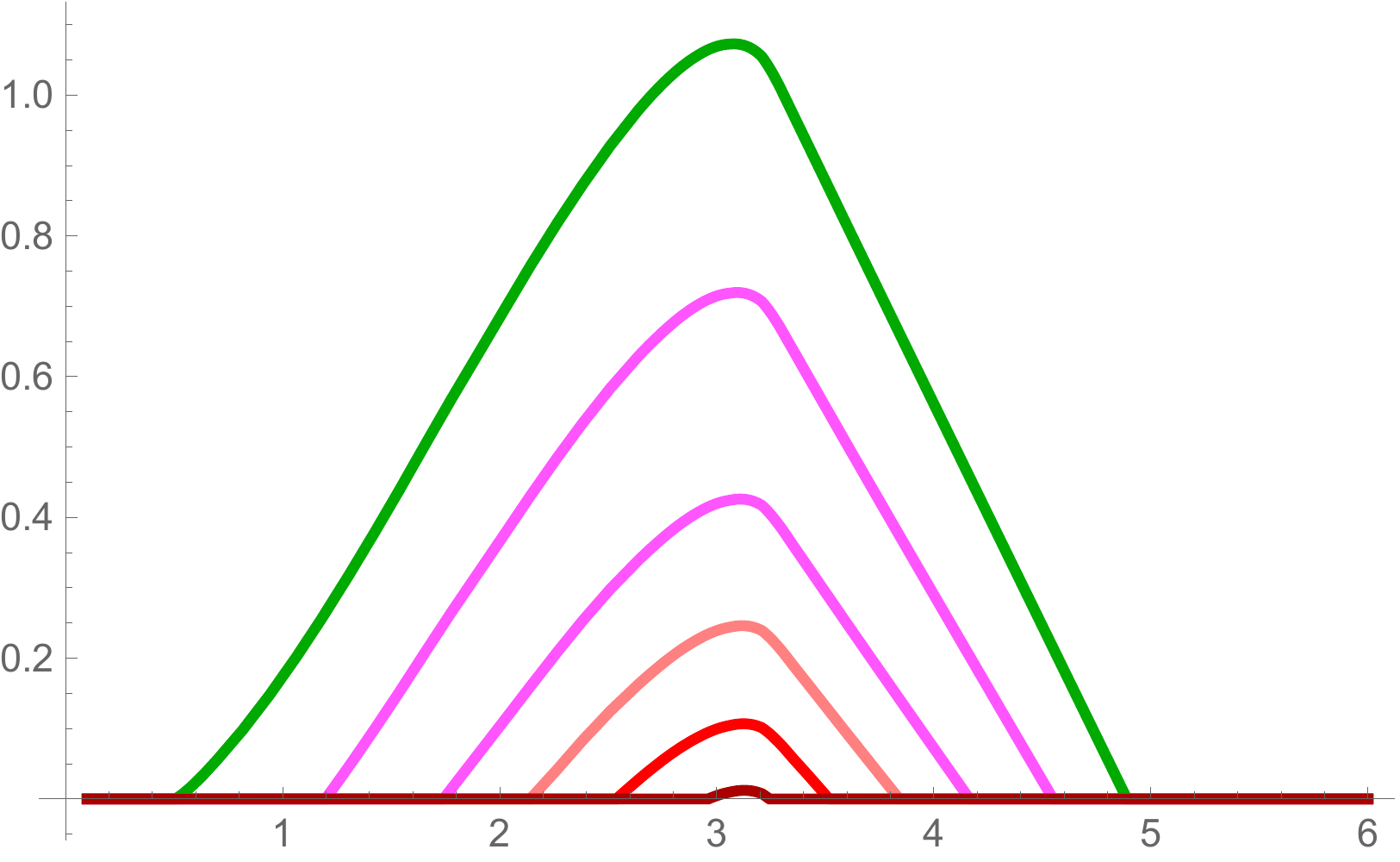}}
\put(20,57){$I$}
\put(233,57){$I$}
\put(75,0){$t$}
\put(330,0){$t$}

\end{picture}
\caption{
 In large plots the scrambling (solid lines)  and the wake up  (dashed lines) times as  functions of the interval length $\ell$ for different initial temperatures parametrized by $z_H$ are presented. The scrambling time corresponds to solid color curves and the wake up time to dashed ones. The left big panel shows $z_H=4.5, 2, 1.7, 1.54, 1.49$ 
and  $z_H =1.47$ (the green line) from the left to right. The right big panel shows $z_H=1.47, 1.37, 1.3, 1.26, 1.23$ and $z_H=1.21$ from the left to right. The green line is the same in the right and left panels. The solid black lines are the thermalization time for the interval size $2l+x$, the dashed ones are the same for the size $l$ and the dotted ones are for the size $x$. The intervals separation is taken to be $x=1$ and the final temperature corresponds to $z_h=1$. 
In the small plots the mutual information time evolution is shown at the same parameters (at by same color) as at large plots.
  }
 \label{fig:tscrl1}
\end{figure}

In Fig.\ref{fig:tscrl1} we plot the dependence of the scrambling and wake times on the interval size, $\ell_1=\ell_2=\ell$, for different initial temperatures
 and the fixed final temperature, $z_h=1$. We see that for $z_H>z_{cr}(z_h)$, $z_{cr}(z_h=1)=1.47$,  there are critical $\ell_{cr,wup}=\ell_{cr,wup}(z_H)$ and $\ell_0(z_H)$, so that only for $\ell_0<\ell\leq\ell_{cr,wup}$ the  wake up time exists. The scrambling time exists for $\ell_0<\ell$. 
 Therefore, for $z_H>z_{cr}(z_h)$  the  bell form of the mutual information evolution
 takes place for the interval $\ell_0<\ell\leq\ell_{cr,wup}$.  For $z_H<z_{cr}(z_h)$, i.e. for high enough initial temperature, it let us remind again, that initial temperauture has to be less than the final one,
 the bell form of the mutual information evolution always exists at  large enough  $\ell$.

\begin{figure}[h!]
\centering
\centering\begin{picture}(185,200)
\put(0,0){
 \includegraphics[scale=0.5]{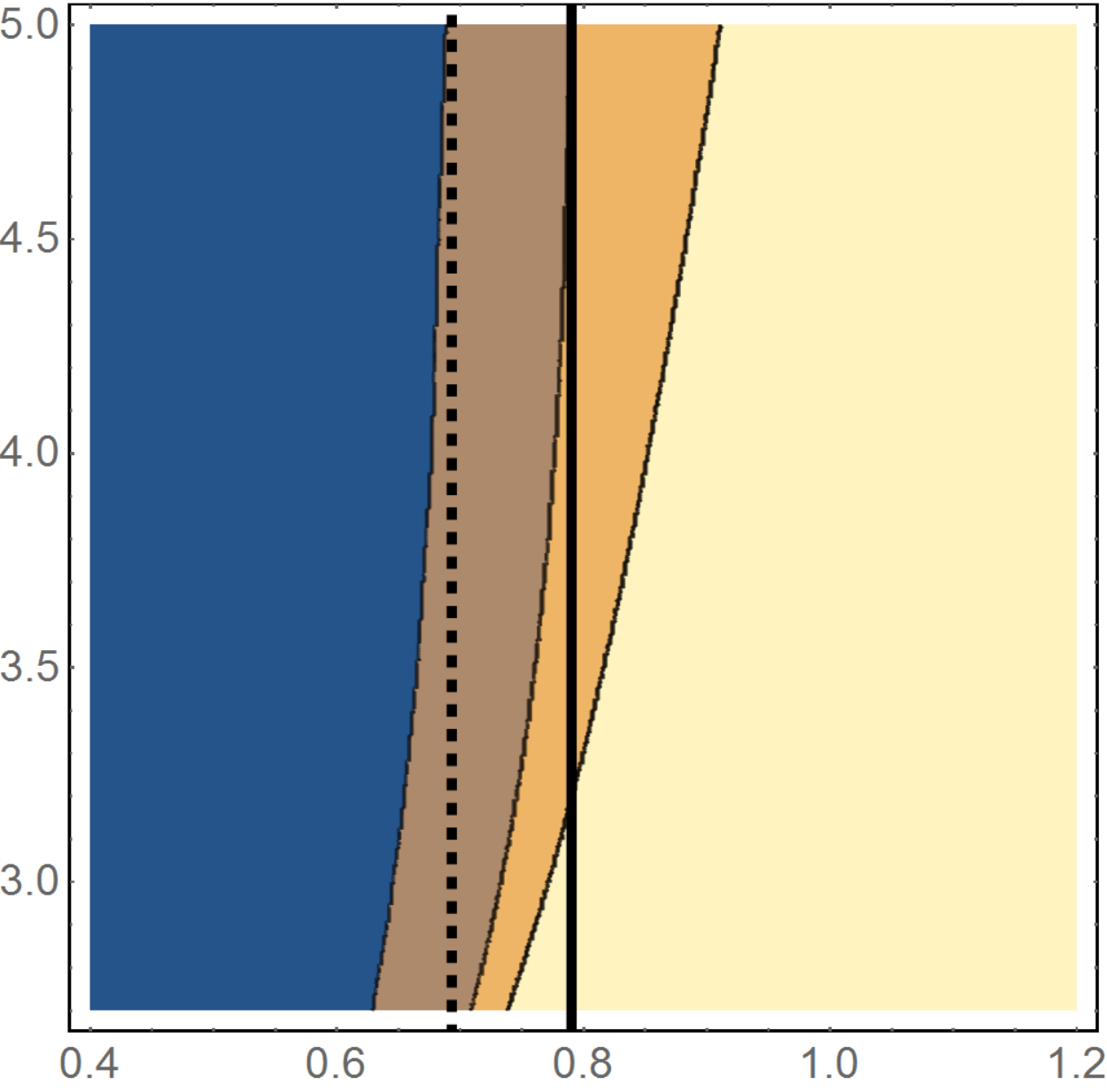}}
  \put(110,-5){$x_{2}$}
    \put(85,-5){$x_{1}$}
\thicklines
         \put(33,22){\line(1,0){90}}
           \put(33,55){\line(1,0){90}}
             \put(33,125){\line(1,0){90}}
              \put(114,10){\line(0,1){117}}
               \put(89,10){\line(0,1){47}}
                 \put(125,20){$\ell_{0}(x_1)$}
                   \put(125,50){$\ell_{wup}(x_{1})$}
                     \put(125,130){$\ell_{0}(x_2)$}
                        \put(70,181){$x_{scr}$}
                          \put(95,181){$x_{wup}$}
                          \end{picture}
 \begin{picture}(185,200)
\put(30,0){
 \includegraphics[scale=0.5]{figs/part-mm.pdf}}
  \put(138,-5){$x_{2}$}
    \put(120,-5){$x_{1}$}
\thicklines
       \put(43,80){\line(1,0){100} }                               
                \put(145,85){$\ell=\ell_{wup}(x_1)$}
                   \put(145,65){$\ell=\ell_0(x_2)$}
                          \put(100,181){$x_{scr}$}
                           \put(130,181){$x_{wup}$}
                           \put(134,10){\line(0,1){67}}
               \put(123,10){\line(0,1){67}}
 \end{picture}\\
 
$\,$
\caption{Zones of different regimes of the mutual information behaviour for different
$\ell$ and $x$ for  $z_H=1.2$ and $z_h=1$.
In the left plot points $(x_{1},\ell_{0}(x_1))$ and $(x_{2},\ell_{0}(x_2))$ are on the border of the orange and yellow zones. In the right plot the points $\ell=\ell_{wup}(x_{1})$ is  are on the border of the orange and yellow zones, meanwhile and 
$(x_2,\ell_{0}(x_2))$  is on the border of the brown and orange zones}
 \label{fig:small-zones}
\end{figure}

 One can see that
our plots Fig.\ref{fig:tscrx1}-Fig.\ref{fig:tscrl1} confirm the zone structure presented in Fig.\ref{fig:wakeup}.
Indeed, let us consider  Fig.\ref{fig:small-zones} that shows a small part of  Fig.\ref{fig:wakeup}. In the right plot 
let us set $x_{1}<z_H\log 2$ fixed. We can always find $\ell_{0}=\ell_0(x_{1})$, so that increasing $\ell>\ell_{0}$ we are in the bell zone till
$\ell<\ell_{wup}(x_{1})$.  However, if  the distance $x$ is taken $x=x_{2}>z_H\log 2$ then the bell zone stars at $\ell_{0}=\ell_0(x_2)$ 
and exists 
till an arbitrary $\ell>\ell_{0}$. If we take $\ell$, then we always can find $x_1$  and $x_2$, so that  $\ell=\ell_{wup}(x_{1})$ and
$\ell=\ell_{0}(x_2)$, i.e. points are on the  borders   of the brown and orange zones and of the orange and yellow zones, respectively, and for any $x$,  $x_1<x<x_2$ the mutual information  evolution has the bell-type form.

  \begin{figure}[h]
\centering
 \includegraphics[scale=0.4]{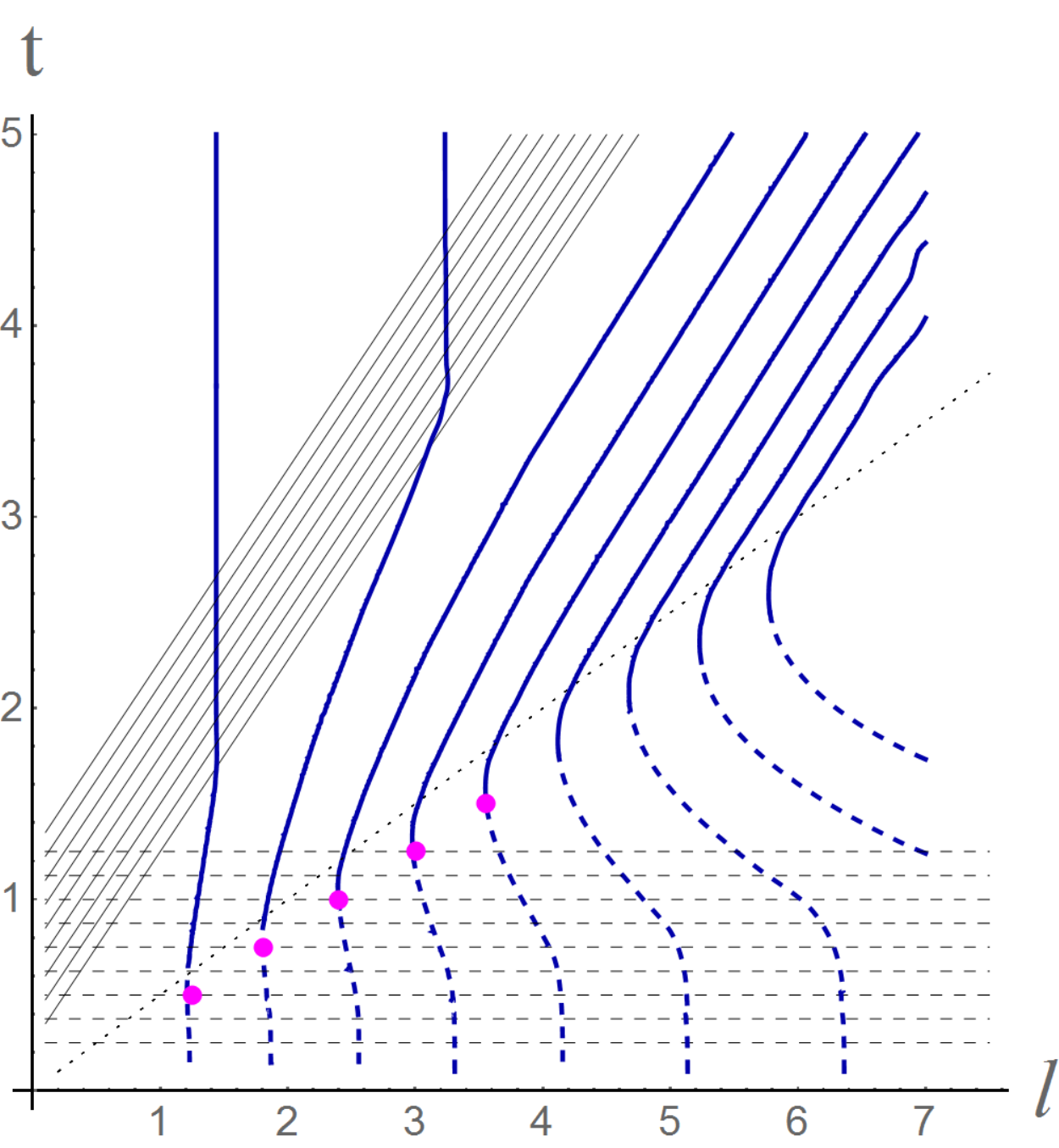}$\,\,\,\,\,\,\,\,$$\,\,\,\,\,\,\,\,$$\,\,\,\,\,\,\,\,$$\,\,\,\,\,\,\,\,$
 \includegraphics[scale=0.4]{figs/pointsandzones.pdf}
  \caption{Points on the border of orange and yellow zones, here  $z_H=4,z_h=1.2$.
 }
 \label{fig:maxI}
\end{figure} 

Fig.\ref{fig:maxI} shows  points that are on the border of the yellow and orange  zones.

\section{Conclusion}
In this paper we have investigated   different aspects of the  holographic mutual information behaviour in the heating process using the AdS/CFT correspondence. The initial state is taken to be thermal, and as a dual background 
describing the heating  process we take the BH-Vaidya metric  in the thin shell approximation. We derive the explicit formulae for the geodesics in this 2+1 dimensional background and then describe the mutual information evolution for the system of two disjoint intervals.
In the consequent work \cite{Ageev:2017wet} using formulae for the evolution of holographic entanglement entropy we obtain the explicit description for different regimes of holographic heating up process.

We have paid the special attention to the bell form of the mutual evolution. This interest is  due to  a crucial role of this type of evolution in the 
 photosynthesis \cite{NUM}. The widest 
 bell-like zone  corresponds to  the symmetric 
configurations and the larger temperature difference between the initial and final temperatures.  This zone exists for  small distances  
$x<z_H\log 2$, only for the particular  interval  sizes, $\ell_{scr}(x)<\ell<\ell_0(x)$ and for x large enough $x>z_H\log 2$,
exists only for large enough interval  sizes, $\ell>\ell_{scr}(x)=\ell(x,z_H)$, where $\ell(x,z_H)$ is given by \eqref{line}.

Notions of the quantum information help us  in the understanding of the
underlying quantum structure of the photosynthesis.
 The appearance  of the bell-type evolution of the mutual information in the thermal states as a result of the global quench 
gives an opportunity of the information exchange between two subsystems even for non-zero temperature
and in the context of   the photosynthesis this can be interpreted as sending the signals from the antenna to the recreation center.
The bell-type evolution under holographic smooth heating of higher dimensional  two strips  has been found numerically in \cite{AV-photo}.
It would be interesting to study the same problem for $3\leq n\leq 7$ regions, especially of  3-dimensional ball forms as well as use
more complicated holographic backgrounds \cite{ABK,1205.1548,1205.2354,1612.00082}.

$$\,$$

\section*{Acknowlegement}
We would like to thank I. V. Volovich and M. A. Khramtsov for useful discussions. This work
was supported by the Russian Science Foundation (grant No.14-11-00687).

     \end{document}